\documentclass[aps,prb,amsmath,amssymb,amsfonts,twocolumn,nofootinbib]{revtex4}

\usepackage{graphicx}
\usepackage{dcolumn}
\usepackage{bm}
\usepackage{amsmath}
\usepackage{amssymb}
\usepackage{color}

\usepackage{adjustbox}
\usepackage{multirow}

\usepackage{soul}

\makeatletter
\renewcommand*\env@matrix[1][*\c@MaxMatrixCols c]{%
 \hskip -\arraycolsep
   \let\@ifnextchar\new@ifnextchar
  \array{#1}}
\makeatother

\newcommand{\const}{\mathrm{const}}

\newcommand{\tr}{\mathrm{tr}}
\newcommand{\im}{\mathrm{Im}}
\newcommand{\re}{\mathrm{Re}}
\newcommand{\be}{\begin{equation}}
\newcommand{\ee}{\end{equation}}
\newcommand{\pp}{\partial}
\newcommand{\bea}{\begin{eqnarray}}
\newcommand{\eea}{\end{eqnarray}}
\newcommand{\la}{\langle}
\newcommand{\ra}{\rangle}
\newcommand{\dg}{\dagger}
\newcommand{\td}{\tilde}
\newcommand{\dgc}{\ddagger}
\newcommand{\G}{\Gamma}

\definecolor{orange}{rgb}{1,0.5,0}

\begin{document}
 \title{
 Power-law approach to steady state in open lattices of non-interacting electrons}

\author{M. V. Medvedyeva}
\affiliation{Institute for Theoretical Physics, Georg-August-Universit{\"a}t G{\"o}ttingen,
Friedrich-Hund-Platz 1, 37077 G{\"o}ttingen, Germany}
\author{S. Kehrein}
\affiliation{Institute for Theoretical Physics, Georg-August-Universit{\"a}t G{\"o}ttingen,
Friedrich-Hund-Platz 1, 37077 G{\"o}ttingen, Germany}

\begin{abstract}
We address the question of how a non-equilibrium steady state (NESS) is reached in the Linbdladian dynamics of an open quantum system. We develop an expansion of the density matrix in terms of the NESS-excitations, each of which has its own (exponential) decay rate. However, when the decay rates tend to zero for many NESS-excitations (the spectral gap of the Liouvillian is closed in the thermodynamic limit), the long-time dynamics of the system  can exhibit a power-law behaviour. 
This relaxation to NESS expectation values is determined by the density of states close to zero spectral gap and the value of the operator in these states. We illustrate this main idea on the example of the lattice of non-interacting fermions coupled to Markovian leads at infinite bias voltage. The current comes towards its NESS value starting from a typical initial state as $\tau^{-3/2}$.  This behaviour is universal and independent of the space dimension.
\end{abstract}
 \maketitle

\section{Introduction}

\subsection{Overview}

Recently, the structure of the non-equilibrium steady states (NESS) in various open quantum systems has attracted a lot of attention\cite{Wiess,ProsHubb,Alb,intLight}. The non-equilibrium steady state in open systems can exhibit properties fundamentally different than in closed systems. For example, depending on the coupling, different order parameters can condense in the system~\cite{sarang}. The system can experience the localization-delocalization transition driven by the coupling to the bath.~\cite{SpinBoson,Sachdev,Werner,Sachdev2,Sachdev3} A separate direction of research  is focused on the construction of states via dissipation\cite{Diehl, Ver,Top,Top2}. 

An important question arises: how quickly is the NESS reached? In the area of molecular junctions this question has already been addressed by solving the evolution equations~\cite{molec, Nov,Rainer}. There, both the system and the reservoirs are fermionic in nature.
Interesting phenomena such as bistability due to the coupling with phonons were investigated~\cite{bist}. 
A precise treatment is possible here, as a molecular junction itself is a small system. 
 Another small system where the relaxation has been studied in detail is the spin in the spin-boson model. There the phase diagram for coherent and incoherent relaxation regimes has been obtained by combining real-time renormalization group~\cite{RTRG} and functional renormalization group~\cite{FRG}, as well as analytical considerations close to non-interacting resonant level model.~\cite{Kashuba} Different regimes of equilibration can be visualized as the positions of poles of the propagator in the complex energy plane.
 It is interesting to consider how fast the equilibration happens in extended systems. 


In  open quantum systems in the thermodynamic limit it is  not yet well understood how the relaxation happens.
For small systems it is possible to solve the system plus reservoirs by some means as mentioned above. For extended systems it is more complicated to use such methods as one would need to treat many more degrees of freedom. Therefore, some approximations have to be made. We assume that the dynamics of the reservoirs is much faster than the dynamics of the system. Then the relaxation dynamics is described by the Lindblad equation~\cite{breuer}, which is a linear differential equation for the density matrix.
 Finite systems will always have a discrete complex spectrum of the evolution operator with at least one exactly zero eigenvalue. The discreteness of the spectrum assures that the smallest decay rate is finite, or in other words that the Liouvillian has a spectral gap.
Therefore, the relaxation is exponential beyond the time-scale determined by the gap. For large systems, the spectral gap can go to zero in the thermodynamic limit~\cite{Prosen10} (the system size goes to infinity). The time dynamics in such systems can show  power-law scaling for correlation functions. Recently, for example, a numerical study has been performed using time-dependent density matrix renormalization group (t-DMRG) and a power-law relaxation to a NESS has been observed.~\cite{Muenc} The system studied is a spin chain coupled to the reservoirs at every site. The  Lindblad-type dissipation is expected to lead to a well-defined steady state, which is an attractor of the time-evolution. The authors connect the power-law decay approach to a NESS to the closing of the spectral gap.    Slower-than-exponential dynamics  has also been observed in a number of works on relaxation in bosonic systems:  algebraic relaxation has been reported in Ref.~\onlinecite{AlgBos} and stretched exponential in Refs.~\onlinecite{bos0,bos1,bos2}. In a fermionic system a stretched exponential coming towards a NESS has been also observed~\cite{SlowFerm}.


In this paper we provide an analytic consideration to investigate why and when the observables have a power-law relaxation in the case when the spectral gap  closes in the thermodynamic limit. We point out that not only the closing of the spectral gap is important to characterize the approach to equilibrium, but also the density of the decay rates close to zero and the values of the matrix elements of the observables for different decay states. We perform our analysis on the example of non-interacting fermions coupled to the leads via Lindblad operators in order to get analytical insights  in the thermodynamic limit. This is important because for any finite system the approach to the NESS eventually becomes exponential. The NESS properties for such a system have been considered before in~Refs.~\onlinecite{Prosen10,Prosen12,Kosov,Buca,Zni10,Pro08,Zni,Medv}, while in this paper we are interested in the approach to the NESS.

Normally the relaxation in dissipative systems is studied by numerical methods.
While many specific examples can be studied numerically, it remains difficult to make general statements for a broad class of systems. Also, it can be a challenge to control the errors and obtain an accurate description.
Here we present an analytical approach. It provides the knowledge on the structure of the density matrix during the time-evolution and therefore applies to all observables. Our technique is independent of the spatial dimension of the system.

 

\subsection{Main idea}

We consider an evolution of  a system coupled to  an environment by the Lindblad operators, $L_{\mu}$:
\bea i\frac{\pp \rho}{\pp \tau} &=& \mathcal{L} \rho ,\label{LindGen}\\
  \mathcal{L} \rho &=& [H,\rho] + i \sum_{\mu}\left(2 L_\mu  \rho L^{\dg}_\mu - \{L^{}_\mu L^{\dg}_\mu,\rho\} \right) ,\label{liouv} \eea
where $H$ is a Hamiltonian of the system and $\tau$ is time.
It is a linear differential equation for the density matrix, unlike  the Schr\"{o}dinger equation which is written for the wave function. The second significant difference with respect to the Schr\"{o}dinger equation is that the Liouvillian, $\mathcal{L}$, is non-Hermitian. The Liouvillian has at least one zero eigenvalue, which determines the NESS solution.~\cite{breuer}   As a consequence, every initial state relaxes to the NESS. The resulting density matrix $\rho_{NESS}$ is time-independent if there is only a single zero in the spectrum of $\mathcal{L}$  (otherwise we could have one of the following situations: (i) if there is a degenerate zero eigenvalue, then the non-equilibrium steady states form a subspace in the state space and the NESS depends on the initial state, (ii) if only the imaginary part of several eigenvalues is zero, but not real, there would be oscillations between different modes of $\mathcal{L}$).
The main question we are going to address is how does the relaxation to the NESS happen.   We consider a case when both the Hamiltonian and the couplings to the baths are time-independent. The system is prepared in an arbitrary (different from the NESS) and the evolution is driven by the Lindblad equation~(\ref{LindGen}),~(\ref{liouv}).  

Assuming that $\mathcal{L}$ is a time-independent operator, the solution of the Eq.~(\ref{LindGen}) can be represented as
\be \rho(\tau)=\rho_{NESS}+\sum_j C_j \upsilon_j\exp(i\Omega_j \tau), \label{rhot0}\ee
where $\rho_{NESS}$ is the kernel of the operator $\mathcal{L}$, the NESS density matrix has unit trace $\tr(\rho_{NESS})=1$ and  $\Omega_j=\omega_j+i\gamma_j$ and $\upsilon_j$ are  (complex) eigenvalues and eigenvectors of $\mathcal{L}$.  The coefficients $C_j$ depend on the initial density matrix $\rho(0)$. 
The structure of the dissipative term in the Lindblad equation ~(\ref{LindGen}),~(\ref{liouv}) preserves the hermiticity of the density matrix and its trace,~\cite{breuer} hence the solution~(\ref{rhot0}) of Eq.~(\ref{LindGen}) can be rewritten as
\begin{subequations}
\label{rhot}
\begin{align}
 \rho(\tau)&=\rho_{NESS}+\sum_{\gamma_j} A_j \rho_j(\tau)\exp(- \gamma_j \tau), \\
  \rho_j (\tau)&= \rho_j^\dagger(\tau) , \tr  \rho_j(\tau) =0,
\end{align}  
\end{subequations} 
where we combine the terms $\upsilon_j\exp(i\Omega_j \tau)$ with the same decay rates $\gamma_j$ into a single expression $\rho_j(\tau)\exp(- \gamma_j \tau)$ which has a  certain decay rate $\gamma_j$ (the dependence on $\tau$ in $\rho_j(\tau)$ comes from the oscillating part $\exp(i \omega_j \tau)$). Expression (\ref{rhot}) is valid for any dissipative system described by the Lindblad equation with time-independent Liouvillian as it is derived from the conservation of the trace and the hermiticity of the density matrix during the time evolution, and these requirements are always satisfied by Lindblad-driven dynamics. The expansion (\ref{rhot}) is made around the NESS, therefore we call {\it $\rho_j(\tau)$ excited states above the NESS} or {\it NESS-excitations}. It is in analogy with the term "excitation" used for the eigenstates above the ground state of the Schr\"{o}dinger equation.
For the Liouvillian the special state is NESS which is characterized by the zero eigenvalue. The NESS-excitations are the eigenstates of the Liouvillian with a certain non-zero decay rate $\gamma_j$. The requirements $\rho_j (\tau)= \rho_j^\dagger(\tau) , \tr  \rho_j(\tau) =0$ make it possible to expand the initial density matrix in the NESS-excitations. 
We cannot call $\rho_j(\tau)$ excited density matrix states as $\rho_j(\tau)$ are traceless, so they are not density matrices.

Let us make a very simple estimate for 
how fast the relaxation to the NESS of some operator happens. We will show later in Secs.~\ref{sec:Formalism}, \ref{sec:Numerics} that the assumptions involved are valid. We assume the decay rates $\gamma_j$ to follow a power-law: $\gamma_j = a j^\beta$, $j=1,2,3,\ldots$; the coefficients $C_j$ are arbitrary and all of similar magnitude; for now let us put all of them equal to some constant $C$. For simplicity we assume that only pure imaginary eigenvalues $\Omega_j$ play a role in the equilibration  (a complex $\Omega_j$ will lead to oscillations, and averaging over them reduces the exponent of the power-law behaviour by one, see expressions below). For computing the expectation value of some observable, we take into account its expectation value for each NESS-excitation $\rho_j(\tau)$. Let us assume these expectation values are proportional to $j^\alpha$. Then the approach to the NESS-value of the observable at hand is estimated as
\begin{subequations} 
\label{est2}
\begin{align}
\la \hat{O}(t) - O_{NESS} \ra &\propto  \sum_j j^\alpha \exp(-j^\beta \tau a) \\
 &\propto  \tau^{-(\alpha + 1)/\beta}, \beta-\alpha >1.
 \end{align}
\end{subequations}  
The power-law approach appears as a natural consequence of the assumptions made above. 

The above sketchy derivation can be reformulated in terms of the density of the decaying states. 
Expectation value of an observable $\hat{O}$ is
$\la \hat{O} \ra = \tr(\rho \hat{O}).$ 
It is often convenient to represent the density matrix in the energy basis (for example, for the system in the thermodynamic equilibrium with environment). In this case the expectation value can be rewritten using the density of states $\nu(\omega)$: 
$\la \hat{O} \ra = \int d\omega \nu(\omega) O(\omega)$. 
For the non-equilibrium situation the energy is complex, which we denoted by $\omega+i\gamma$. The decay rate of the operator approaching the steady state value can be represented as an integral over the density of the decay states $d\nu(\gamma)$:
\be \label{FDT} \la \hat{O}(t) - O_{NESS} \ra = \int d\gamma \nu(\gamma) O(\gamma)\exp(-\gamma \tau).\ee
As in the previous derivation we have neglected the oscillations in time as they lead to a faster decay rate. In terms of the previously introduced exponents $\alpha$ and $\beta$ the density of states at $\gamma\rightarrow 0$ is $\nu(\gamma)\propto \gamma^{1/\alpha-1}$, and $O(\gamma)\propto \gamma^{\beta/\alpha}$.    

The main goal of our paper is to derive in a more precise way the expressions~(\ref{rhot}, \ref{est2}) sketched above for the case of the non-interacting fermionic chain coupled at its ends to the Markovian reservoirs. 
We will also discuss the role of the coefficients $C_j$ which depend on the initial state of the equilibration process. Numerically we find that indeed for most states with a sparse density matrix in position space the time-dependence of the equilibration is in a good agreement with an assumption of equal coefficients $C_j$ for all $j$. We illustrate the formalism by computing the time-dependence of the  expectation value of the current. 

\section{Formalism}
\label{sec:Formalism}

\subsection{Diagonalization of the Liouvillian}

Here we consider a chain of non-interacting fermions of length $N$ described by the tight-binding Hamiltonian:
\be \hat{H}=  \sum_{i} t \left( a^\dg_i a_{i+1}+ h.c \right)+ \sum_i \mu_i a^\dg_i a_i \label{ham},\ee
where $t$ is the hopping matrix element between the neighbouring sites and $\mu_i$ is an on-site potential. 
Evolution of a system coupled to the memoryless bath is described by the Lindblad equation ~(\ref{LindGen}),~(\ref{liouv}). 
The chain of fermions is coupled to the source and the drain at infinite bias voltage~\cite{Gurv1,Gurv2} at the ends of the chain: $L_1^{(i)}=\sqrt{\Gamma_1^{(i)}}a_1^\dg$, $L^{(o)}_N=\sqrt{\Gamma_N^{(o)}}a_N$,  where the superscript $(i)$ stands for the incoming electrons, and the superscript $(o)$ for the outgoing electrons from the lattice to the reserviors.  
 In further equations we measure the dissipation rates $\Gamma$ in the units of hopping matrix element $t$, and thus put $t=1$.

The solution of the Lindblad equations for non-interacting fermions is notably simplified in the super-fermionic representation (for more details see Refs.~\onlinecite{Kosov, Medv}).  In this method two types of operators are introduced, acting on the density matrix from the left and from the right $\{\td{a},\td{a}^\dg\}$. Those acting from the left are "ordinary" operators and those acting from the right are denoted by a tilde.  
In this representation, instead of solving a differential equation for the evolution of the $2^N\times 2^N$  density matrix, the calculations are done with the $2N\times 2N$ matrices. 

The Liouvillian for non-interacting fermions in the super-fermionic representation becomes quadratic after performing the particle-hole transformation~\cite{Kosov,Medv}. The Liouvillian becomes diagonal in the transformed basis, which we denote as $\{f,f^\dgc,\td{f},\td{f}^\dgc\}$:
\be \mathcal{L}_f = \sum_i \lambda_i f_i^\dgc f_i - \sum_i \lambda_i^* \td{f}_i^\dgc \td{f}_i. \label{LDiagonal}\ee
where $f$ and $\tilde{f}$ act on the vectors from the new state space and $f^\dgc$ and $\tilde{f}^\dgc$ act on the vectors from the corresponding dual space. Let us note that the operators $f$ and $f^\dgc$ are not Hermitian conjugate, but only dual to each other (they obey the fermionic anticommutation rules). That is why we use the symbol ${}^{\dgc}$ instead of ordinary ${}^{\dg}$ for the dual operators.
The values $\lambda_i$ are computed as the eigenvalues of the matrix $\mathcal{N}$ that we introduced in Ref.~\onlinecite{Medv} (for a fast recapitulation see Appendix~\ref{sec:LTransformation}). 
The operators $\{f,f^\dgc,\td{f},\td{f}^\dgc\}$ are linear combinations of the operators $\{a,a^\dg,\td{a},\td{a}^\dg\}$:
\begin{subequations}
\label{afTransform}
\begin{align}
a_m^\dg = \sum_{k_1}C^{(1)}_{mk_1}f^\dgc_{k1} + C^{(2)}_{mk_1}\td{f}_{k1}, \label{eq1}\\
a_m = \sum_{k_1}A^{(1)}_{mk_1}f_{k1} + A^{(2)}_{mk_1}\td{f}^\dgc_{k1}. \label{eq2}
\end{align}
\end{subequations}
The coefficient matrices $C^{(1,2)}$ and $A^{(1,2)}$ are connected to the matrix of the eigenvalues $P$ of the matrix $\mathcal{M}$ (see Ref.~\onlinecite{Medv}).
The details about the transformation from the $\{a\}-$basis to the $\{f\}-$ basis are summarized in Appendix~\ref{sec:LTransformation}.
In the $\{f\}$-basis NESS is the vacuum state, therefore observables in the NESS can be computed directly. 

 What are the $\lambda$'s which enter the diagonalized form of the Liouvillian in~Eq.~(\ref{LDiagonal})?
In the absence of the coupling to the environment they are $\lambda_j^{(0)}=2\cos \frac{\pi j}{N+1}$. It is a dispersion relation of the tight-binding model with Hamiltonian (\ref{ham}), $U_i=0$, $t=\const.\rightarrow 1$. Upon turning on the coupling to the environment,  $\lambda_j$ gains an imaginary part. It indicates the decay rate of one-particle modes towards the NESS.
The real part of $\lambda_j$ can still be viewed as a determining the dispersion relation in the dissipative tight-binding model.

\subsection{Time evolution}

 We are interested in the time evolution of the density matrix and the expectation values of the observables.
The expectation value of an observable $\hat{O}$ in the super-fermionic formalism~\cite{Pro08} is 
\be \label{ExpValue}
\tr (\rho(\tau) \hat{O})=
{}_{f^\dgc,\td{f}^\dgc} \la 0| \hat{O}_f \rho_f(\tau) | 0 \ra_{f,\td{f}},\ee
where by $\rho_f(\tau)$ we have denoted the density matrix in the $f$-basis. It can be viewed as a linear combination of the NESS and the NESS-excitations. The density matrix of the NESS is  $\rho_{NESS}={}_f|0\ra \la 0|_{f}$, the vacuum in the $\{f\}$-basis. The general density matrix can be expressed as a polynomial $\mathcal{P}$ in the operators $f^\dgc,\td{f}^\dgc$ acting on $\rho_{NESS}$: $\mathcal{P}(f^\dgc,\td{f}^\dgc)\rho_{NESS}$. The polynomial $\mathcal{P}$ should have the structure assuring that the resulting density matrix is Hermitian, positive definite and has unit trace. The  unit trace condition is trivially satisfied as the vacuum expectation value of the creation operators is zero, only $\rho_{NESS}$ contributes to the trace of the whole density matrix, for which the trace is one from the normalization of the vacuum~\cite{Medv}.  We construct the terms of $\mathcal{P}$ in such a way that their time evolution has the same form as the time evolution of $\rho_j$ in (\ref{rhot}). The coefficients in front of the terms with the same decay rates $\gamma_j$ are determined  by requiring the hermicitity of the resulting sum with a certain decay rate. 
The full spectrum of the Liouvillian $\Omega_k$ can be recovered from the $2N\times 2N$ representation of the problem as the sums of various combinations of the $\lambda_i$ and $\lambda_j^*$ parameters. Even more, using the definition of the operators $\{f,f^\dgc,\td{f},\td{f}^\dgc\}$ we can write the density matrices $\rho_j(\tau)$, which appear as the combinations of the excited states over the vacuum in the $f$-representation of the problem.  The corresponding density matrices can be derived by induction as pointed out in the Appendix~\ref{sec:hierarchy}. 
In the Appendix~\ref{sec:example} we give a systematic rewriting of the excited states for the two-site chain taking into account the phase factors. 
Positive definiteness of the density matrix is determined by the coefficients $A_j$ in~(\ref{rhot}).  Positive definiteness can be established only by calculating the eigenvalues of the whole density matrix.

\begin{figure}[tb!]
\begin{center}
A\includegraphics[width=0.85\linewidth]{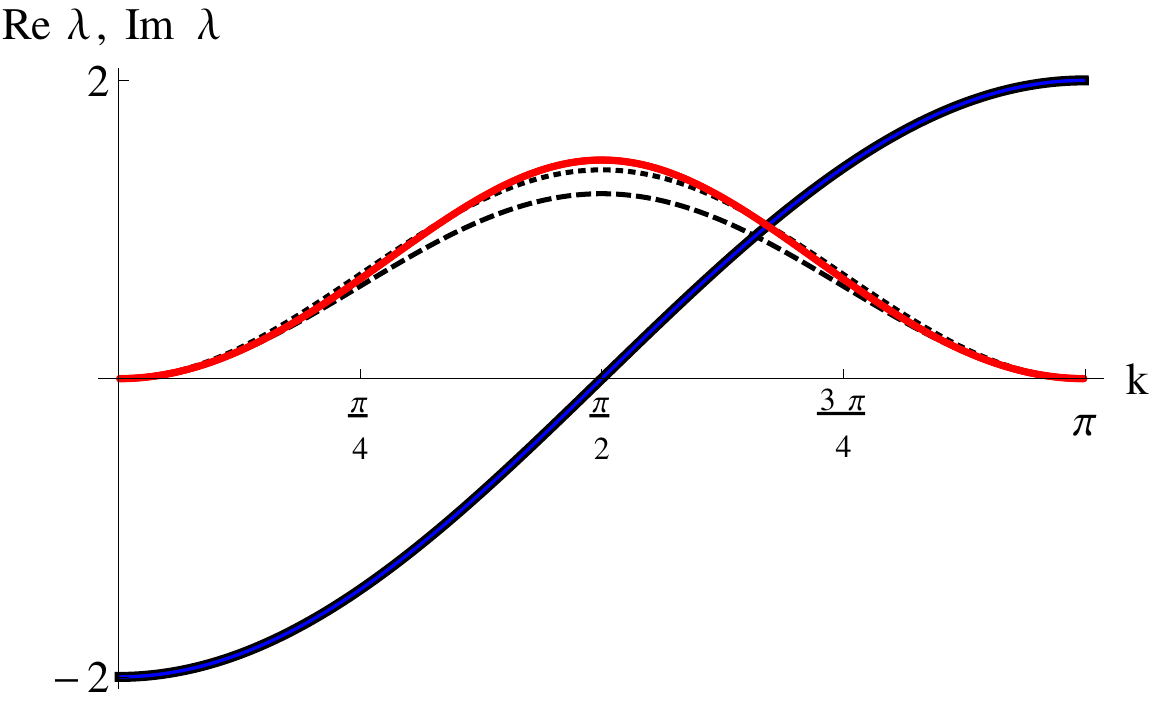}
B\includegraphics[width=0.85\linewidth]{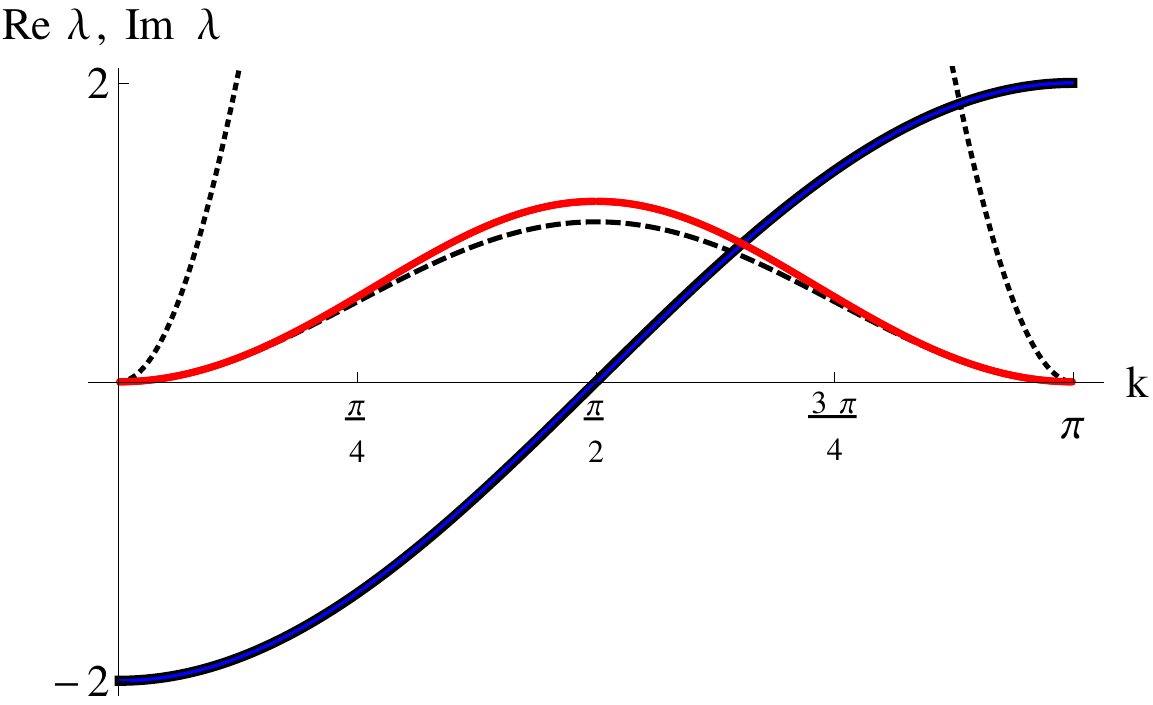}
\caption{\label{dispersion}
The real (blue thick line) and imaginary (red thick line) part of the dispersion relation of the Liouvillian for couplings to environment $\Gamma_1^{(i)},\Gamma_N^{(o)}=(0.3,0.4)$ (A), (3,4) (B).
Imaginary part: the red line is the exact imaginary part of the eigenvalues of the matrix $\mathcal{N}$, the dotted black line is the first order correction (\ref{dlambda1}), the dashed black line is the $1/N$ correction ~(\ref{deltaL2}).
For the real part the perturbative calculations are indistinguishable from the exact ones.   
The chain length is $N=800$ sites. The imaginary part is multiplied by $N+1$.
}\end{center}
\end{figure}

The expression for the $\lambda_j$ in the thermodynamic limit $N\rightarrow \infty$ in the first order in $1/N$ is:
\bea  \lambda_j &=& \lambda_j^{(0)}+  \delta \lambda_j + O(1/N^2),\label{lambda0} \\ 
\delta \lambda_j &=& -i \frac{2 \sin^2\psi_j}{N}\left(\frac{\Gamma_1^{(i)}}{1+\Gamma_1^{2(i)}}+\frac{\Gamma_N^{(o)}}{1+\Gamma_N^{2(o)}} \right) - \\
 &-&\frac{2\sin^2\psi_j}{N}\left(\frac{\Gamma_1^{2(i)}}{1+\Gamma_1^{2(i)}}+\frac{\Gamma_N^{2(o)}}{1+\Gamma_N^{2(o)}} \right), 1\ll N \label{deltaL2}\eea
with $\psi_j=\tfrac{\pi j}{N+1}$.
The correction $\delta \lambda_j$ is obtained directly from the characteristic polynomial of $\mathcal{N}$, see Appendix~\ref{sec:EigenValues}. 
The decay is the slowest for the modes with the lowest energies and with the highest energies, Fig.~\ref{dispersion}. If there is a coupling to the environment such as phonons in a condensed matter system or decay of the excited states of  atoms in  optical lattices, then the high-energy modes can be damped by these processes. We do not include this effect in our model. 

\subsection{Time evolution of the current}
Let us consider the time evolution of a quadratic observable. 
An observable which is quadratic in $\{a\}$-basis remains "quadratic" also in the $\{f\}$-basis. By "quadratic" we mean any contributions of exactly two operators, for example $f^\dgc \td{f}^\dgc$ is also "quadratic" as it contains only two operators. 

Only the NESS-excitations involving two operators $f_j^\dgc,\td{f_k}^\dgc$ contribute to the expectation value of a "quadratic" operator, as follows from the expression for the expectation value, Eq.~(\ref{ExpValue}). There are three types of such NESS-excitations, Table~\ref{tab:3types}.
We enumerate the NESS-excitations as $\rho^{(h)}_{m,\{k_1\ldots k_h\}}$, where the superscript $(h)$ stands for the order of the polynomial in $f^\dgc, \td{f}^\dgc$, the set $\{k_1\ldots k_h\}$ denotes the operators which enter in the polynomial and the decay rate for each excitation determined by the operators with indices $\{k_1\ldots k_h\}$ is $\sum_{k \in \{k_1\ldots k_h\}} {\im \lambda_{k}}$. If several different polynomial in $f_j^\dgc,\td{f_k}^\dgc$ correspond the the same decay rate, then we distinguish the polynomials by the index $m$. Such NESS-excitations are different physically, as they have different oscillation frequencies.

\begin{widetext}
\begin {table*}[ht]
\caption {Three types of quadratic NESS-excitations. $p,s,q$ are pre-factors which make sure that the excited density matrix is Hermitian.} \label{tab:3types} 
\begin{center}
\begin{tabular}{|c|c|c|c}
\hline
decay rate & decay in time & excited state\\ \hline
$2\im \lambda_j$ & $\exp(-2\im \lambda_j \tau)$ & $ \rho^{(2)}_{jj} = p f_j^\dgc \td{f_j}^\dgc,$ \\
$\im \lambda_j+\im \lambda_k$ & $\exp(-(\im \lambda_j+\im \lambda_k) \tau)$ &
$\rho^{(2)}_{1,jk}= s_{1,jk}e^{i (\re \lambda_j+\re \lambda_k) \tau }f_j^\dgc f_k^\dgc  + s_{2,jk} e^{-i (\re \lambda_j+\re \lambda_k) \tau} \td{f}_j^\dgc \td{f}_k^\dgc,$ \\
$\im \lambda_j+\im \lambda_k$ &
$\exp(-(\im \lambda_j+\im \lambda_k) \tau)$ &
 $\rho^{(2)}_{2,jk}=q_{1,jk}e^{i (\re \lambda_j-\re \lambda_k) \tau}f_j^\dgc \td{f_k}^\dgc +q_{2,jk} e^{-i (\re \lambda_j-\re \lambda_k) \tau}f_k^\dgc \td{f_j}^\dgc$\\
 \hline
\end{tabular}
\end{center}
\end{table*}
\end{widetext}

The long-time behaviour is determined by the $\lambda_k$'s with the smallest imaginary part.  In the thermodynamic limit they scale as $k^2/N^3, k=1,2,\ldots,k_{e}$, with $k_e\ll N$ according to Eq.~(\ref{deltaL2}). It is the result which was announced in the Introduction. 

Consider now a specific quadratic operator -- the current operator between the sites $k$ and $k+1$:
\be \hat{j}_k=-i (a^\dg_{k+1}a_k-a^\dg_{k}a_{k+1}).\ee

The evolution of the operators in the $\{f\}-$basis is simple as the evolution operator, $\mathcal{L}$, is diagonal, Eq.~(\ref{LDiagonal}):
\bea \label{Fevolution} f_j(\tau)&=& e^{-i \lambda_j \tau} f_j,~~~ f^\dgc_j(\tau)= e^{i \lambda_j \tau}f^\dgc_j, \\
\td{f}_j(\tau)&=& e^{i \lambda^*_j \tau} \td{f}_j,~~~ \td{f}^\dgc_j(\tau)= e^{-i \lambda^*_j \tau}\td{f}^\dgc_j.
\eea
The time-dependent expectation value of the current is expressed in the basis $\{f\}$ taking into account the linear relation between the basis $\{f\}$ and $\{a\}$
and time evolution of the $\{f\}$-basis:
\begin{widetext}
\be \label{TDexpvalue}\la a^\dg_m a_n \ra (\tau) = {}_f\la 0 | 
(\sum_{k_1}C^{(1)}_{mk_1}f^\dgc_{k_1} + C^{(2)}_{mk_1}\td{f}_{k_1})
(\sum_{k_2}A^{(1)}_{nk_2}f_{k_2} + A^{(2)}_{nk_2}\td{f}^\dgc_{k_2})
 \rho_f (\tau)|0\ra_f\ee
 \end{widetext}
The NESS expectation value is given by 
$\sum_{k_1}C^{(2)}_{mk_1} A^{(2)}_{nk_1}$. 
The NESS-excitations which contribute to the expectation value are $\rho_{jj}^{(2)}$ and $\rho_{2,jk}^{(2)}$. 
The expectation value for $\rho_{jj}^{(2)}$ is 
$ C^{(1)}_{m j} A^{(2)}_{nj}$  
and for the $\rho_{2,jk}^{(2)}$ it is
$C^{(1)}_{m k} A^{(2)}_{nj}$.
We can estimate these contributions in the thermodynamic limit knowing the perturbative corrections to the eigenvectors of~$\mathcal{M}$, see Appendices~\ref{sec:EigenVectors},~\ref{subsec:Invertion}. 
Then the corrections to the NESS value of the current close to the ends of the chain are  
\begin{subequations}
\begin{align}
\delta \hat{j}_m[\rho_{kk}^{(2)}] &=\tr \hat{j}_m\rho_{kk}^{(2)} \propto g_{1,m}\left(\frac{\Gamma_{N}^{(o)}}{1+\Gamma_{N}^{2(o)}},\frac{\Gamma_{1}^{(i)}}{1+\Gamma_{1}^{2(i)}} \right) \frac{k^2}{N^{3}}  , \label{cor2}\\ 
  \delta \hat{j}_m[\rho_{2,kl}^{(2)}] &=\tr \hat{j}_m\rho_{2,kl}^{(2)}  \propto 
g_{1,m}\left(\frac{\Gamma_{N}^{(o)}}{1+\Gamma_{N}^{2(o)}},\frac{\Gamma_{1}^{(i)}}{1+\Gamma_{1}^{2(i)}} \right) 
  \frac{ k l}{N^{3}}, \label{cor22}
\end{align}
\end{subequations}
where the function $g^{m}_{1}\left(\frac{\Gamma_{N}^{(o)}}{1+\Gamma_{N}^{2(o)}},\frac{\Gamma_{1}^{(i)}}{1+\Gamma_{1}^{2(i)}} \right)$ approaches  $\frac{\Gamma_{N}^{(o)}}{1+\Gamma_{N}^{2(o)}}$ close to the drain and $\frac{\Gamma_{1}^{(i)}}{1+\Gamma_{1}^{2(i)}} $ close to the source. 
We see that the current equilibration as well as the current itself~\cite{Medv, Groth} experiences the quantum Zeno effect. The large values of the coupling to the reservoirs acts as the constant measurement at the ends of the system, leading to the localization of the state at the ends of the chain and consequently decreasing the current. 

Now we can estimate the time-dependence of the approach to the NESS by summing over different NESS-excitations, Eq.~(\ref{rhot}), Table~\ref{tab:3types}.
The contribution coming from  $\rho_j^{(2)}$ is 
\be \label{EqCurrent}  \sum_k \delta \hat{j}_m[\rho_{k}^{(2)}] e^{-2\im\lambda_k \tau} \propto \left(\frac{N+1}{\tau}\right)^{3/2},\ee
and the contribution from  $\rho_{2,jk}^{(2)}$ after averaging over fast oscillations is
\be \label{EqCurrent2} \sum_{l,k} \delta \hat{j}_m[\rho_{2,lk}^{(2)}] e^{-(\im\lambda_k + \im\lambda_l) \tau} \propto \left(\frac{N+1}{\tau}\right)^{3}.
\ee
Therefore, the main contribution to the decay comes from the non-oscillating NESS-excitations. 
We have approximated the sum by  an integral and taken the upper limit to  infinity, as the tail of the function contributes less then the part close to zero. 


 It is possible to investigate the evolution of any operator in a similar way as we have discussed for the current. 
First, one represents the operator in the $\{f\}$-basis, as in Eq.~(\ref{TDexpvalue}). Second, one figures out  what kind of NESS-excitations play a role, depending on the representation of the operator in the $\{f\}$-basis. Generically we can tell that the number of the operators in the NESS-excitation is not larger then the number of the operators in the observable. Knowing the type of NESS-excitations involved, we see which coefficients from the matrices $A^{(1,2)}$ and $C^{(1,2)}$ contribute and perform the summation over the repeating indices taking into account time dependence. 
Therefore, the long-time behaviour of the observable can differ depending on the density of the decay rates $\nu(\gamma)$ and the expectation value $O(\gamma)$ on the corresponding NESS-excitations, see Eq.~(\ref{FDT}).

\begin{figure}[tb]
\begin{center}
\includegraphics[width=0.85\linewidth]{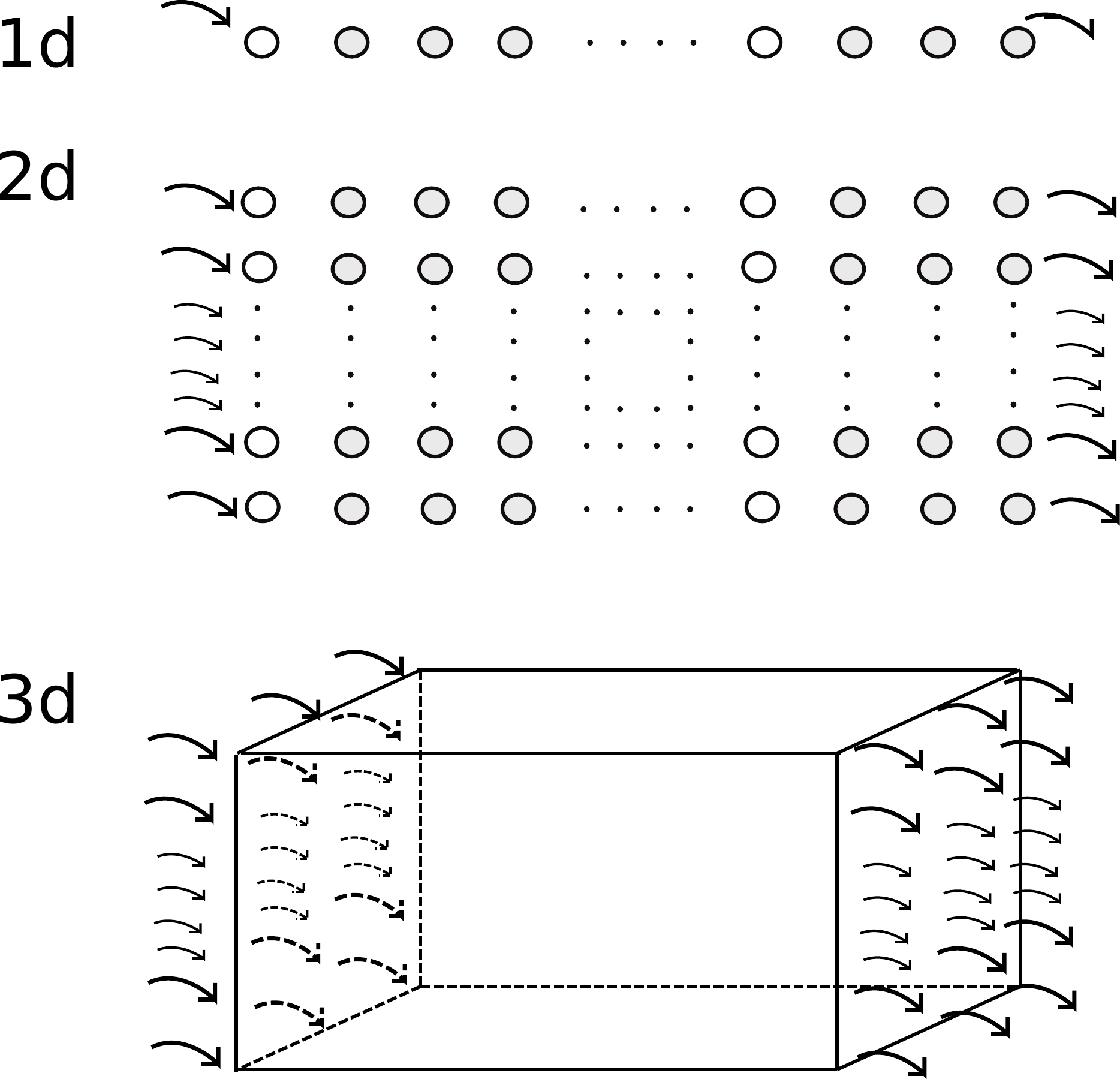}
\caption{\label{nd3}
 The geometry of the current flow through the sample for different dimensions.
}\end{center}
\end{figure}

\subsection{Influence of interactions}

 Let us consider an interacting one-dimensional model with nearest-neighbor interaction of the form $U a_i^\dg a_i a_{i+1}^\dg a_{i+1}$ and remove the interacting part perturbatively by applying a unitary transformation to the Hamiltonian.\cite{Kehrein} 
The weak interactions in the lattice model  lead to a renormalization of the quasiparticle dispersion. The Lindblad operators in the initial, non-renormalized problem, are connected to creation/annihilation operators. They change under the applied unitary transformation to the sum of initial creation/annihilation operators $\{a,\td{a},a^\dg,\td{a}^\dg\}$ with prefactors proportional to the interaction strength. The couplings to the bath (Lindblad operators) are thus modified. The structure of the decay rates is analogous to~(\ref{deltaL2}). But  the decay rates are now proportional to the sum of $\tfrac{j^2}{N^3} \left(\frac{\Gamma_1^{(i)}}{1+\Gamma_1^{2(i)}}+\frac{\Gamma_N^{(o)}}{1+\Gamma_N^{2(o)}} \right)$ and $\tfrac{\mathcal{U} j^2}{N^2}$ (the second term comes from the renormalization of the Lindblad operators, which now become non-local). Therefore, the time-scale when the power-law decay finishes is determined by the maximum of these two contributions. 
The value of the current on the NESS-excitations has the same decay rate as previously. Therefore, weak interaction $\mathcal{U}$ in the system does not change the power-law decay, but might only influence the time-scale on which the power-law decay to the NESS happens.

\subsection{Different dimensions}

One can perform a similar analysis of the decay rates of the NESS-excitations and the value of the operator on these NESS-excitations for uniform stripes in any dimension $d$ connected uniformly to the source and the drain at the ends via the surfaces of the dimension $d-1$, Fig.~\ref{nd3},  and obtain the same power-law decay rate as in one dimension.   
We do not provide an analytical calculation here but give a numerical support of this statement for two-dimensional geometry in the next section. 


\section{Numerical simulation}
\label{sec:Numerics}

In the previous section we have seen that the decay of the initial state towards the NESS can be estimated by summing over the NESS-excitations $\rho_k^{(2)}$. We have assumed the coefficients in front of these states are approximately equal when performing the summation. This is only true for some initial states and one even can argue that it is possible to create a state which will decay much faster or much slower. That is why we check how valid the above treatment is  
for the initial states given in the $\{a\}$-basis. The transformation between the bases, although linear, produces dense state operators in the $\{f\}$-basis from the sparse density matrices in the $\{a\}$-basis  and vice versa. Here we demonstrate the statements from the previous section on the example of the sparse initial density matrices in the $\{a\}$-basis (no matter whether pure or mixed), and show that the power-law of the decay towards the NESS is indeed the same as predicted above.

\subsection{Evolution via diagonalization}

To conduct the evolution numerically we represent the initial density matrix in the $\{a\}$-basis using the creation operators $a^\dg$ and $\td{a}^\dg$, then transform the state to the diagonal basis of the Liouvillian, the basis $\{f\}$, perform the evolution on this basis, Eqs.~(\ref{Fevolution}), and make an inverse transformation to the basis $\{a\}$. 

\begin{figure}[htb]
\begin{center}
\includegraphics[width=0.85\linewidth]{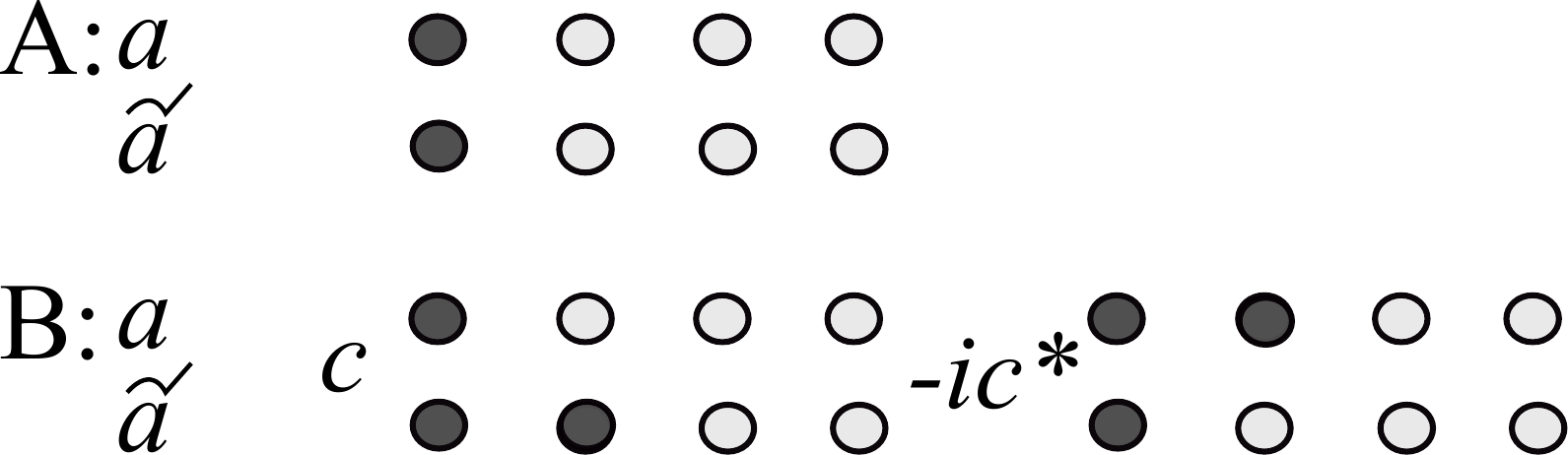}
\caption{\label{densitymatrices}
Schematic representation of the A: diagonal, Eq.~(\ref{diag}) and B: off-diagonal, Eq.~(\ref{nondiag}), parts of the density matrices. Circles represent the structure of the tight-binding model. Dark circles correspond to applied creation operators. 
}\end{center}
\end{figure}

The initial density matrix is represented as the sum over the states of the chain of operators  $a$ and $\td{a}$:
\be \rho=\rho_{diag}+\rho_{non-diag}.\ee
\begin{widetext}
The diagonal part of the density matrix correspond to the operator expressions 
\be\label{diag} \rho_{diag}=\sum_{k=0}^{N}\sum_{\{i_1i_2\ldots i_k \}} \gamma_{\{i_1i_2\ldots i_k \}} a^\dg_{i_1} a^\dg_{i_2} \ldots a^\dg_{i_k} \td{a}^\dg_{i_1} \td{a}^\dg_{i_2} \ldots \td{a}^\dg_{i_k} | 0\ldots00\ldots0\ra_{\{a,\td{a}\}}, \ee
where the summation is performed over all distinct sets of $k$ non-repeating indices $\{i_1i_2\ldots i_k\}$, $0\le k \le N$, $i_j \in (0,\ldots,N),~j\in(1,\ldots,k)$, the state $| 0\ldots00\ldots0\ra$ (or, without using the tilde notation $| 0\ldots0\ra \la0\ldots0|$) is the vacuum for the operators $\{a,\td{a}\}$. In this way many-particle states are taken into account as we are interested in an open quantum system, where the number of particles is not conserved. 
The non-diagonal density matrices are represented as the superposition of two states:
\bea \rho_{non-diag}=\sum_{k,m}\sum_{\substack { \{i_1i_2\ldots i_k \},\{j_1j_2\ldots j_m \}, \\
\{i_1i_2\ldots i_k \}\ne\{j_1j_2\ldots j_m \}}
} & \biggl[ \alpha_{\{i_1i_2\ldots i_k \},\{j_1j_2\ldots j_m \}} a^\dg_{i_1} a^\dg_{i_2} \ldots a^\dg_{i_k} \td{a}^\dg_{j_1} \td{a}^\dg_{j_2} \ldots \td{a}^\dg_{j_m}| 0\ldots00\ldots0\ra_{\{a,\td{a}\}} & + \nonumber\\ 
&\beta_{\{i_1i_2\ldots i_k \},\{j_1j_2\ldots j_m \}} a^\dg_{j_1} a^\dg_{j_2} \ldots a^\dg_{j_m} \td{a}^\dg_{i_1} \td{a}^\dg_{i_2} \ldots \td{a}^\dg_{i_k}  | 0\ldots 00\ldots 0\ra_{\{a,\td{a}\}} \biggr]&, \label{nondiag}\eea
\end{widetext}
where $\alpha$ and $\beta$ stand here to ensure the hermiticity of the resulting density matrix. 
Both types of the density matrices, diagonal and non-diagonal, are schematically represented in Fig.~\ref{densitymatrices}.
Therefore, to evolve the whole density matrix one needs to determine $2^N$ diagonal coefficients $\gamma_{\{i_1i_2\ldots i_k \}}$ and $2^{N-1}(2^N-1)$ the off-diagonal coefficients $\alpha_{\{i_1i_2\ldots i_k \},\{j_1j_2\ldots j_m \}}$. The number of initial coefficients is significantly decreased when one takes into account an operator under considerations. For a quadratic operator one needs to take into account only diagonal elements, and those off-diagonal elements, for which the sets of the indices  $\{i_1i_2\ldots i_k \}$ and $\{j_1j_2\ldots j_m \}$ differ only in two positions. 

Using the representation of the initial density matrix as sum over states of the chains, Fig.~\ref{densitymatrices}, we can perform the evolution by transforming to the diagonal basis $\{f\}$, evolving following Eqs.~(\ref{Fevolution}) and transforming back to the basis $\{a\}$. 

The transformation between the bases $\{a\}$ and $\{f\}$ is highly non-local, Eqs.~(\ref{afTransform}). Therefore, to obtain generic coefficients in the evolution of the density matrix in the $\{f\}$-basis it is sufficient to investigate the initial states containing only a small number of diagonal and non-diagonal elements (some fluctuation) in the $\{a\}$-basis. To get a physical understanding of the processes in the system we also consider the initial density matrices for completely empty chain (for example, if at the initial moment the chain was connected only to the drain) and for completely filled (initially the chain was connected only to the source). 
The case of the density matrices containing only a small number of diagonal and non-diagonal elements corresponds to an empty chain with some fluctuation.

The positive definiteness of the initial density matrix for the arbitrary density matrices is checked explicitely via diagonalization, as we choose a sparse density matrix.

There are fast-oscillating contributions coming from the matrices $\rho_{jk}^{(2)}$ in the $\{f\}$-basis when we consider an evolution of a quadratic operator. Therefore, while performing the time evolution numerically we need to have a time resolution of these fast oscillations. Having such a time-resolution we average over them. The fits are performed after averaging over the fast oscillations.

\subsection{Results}

\begin{figure}[tb!]
\begin{center}
A\includegraphics[width=0.85\linewidth]{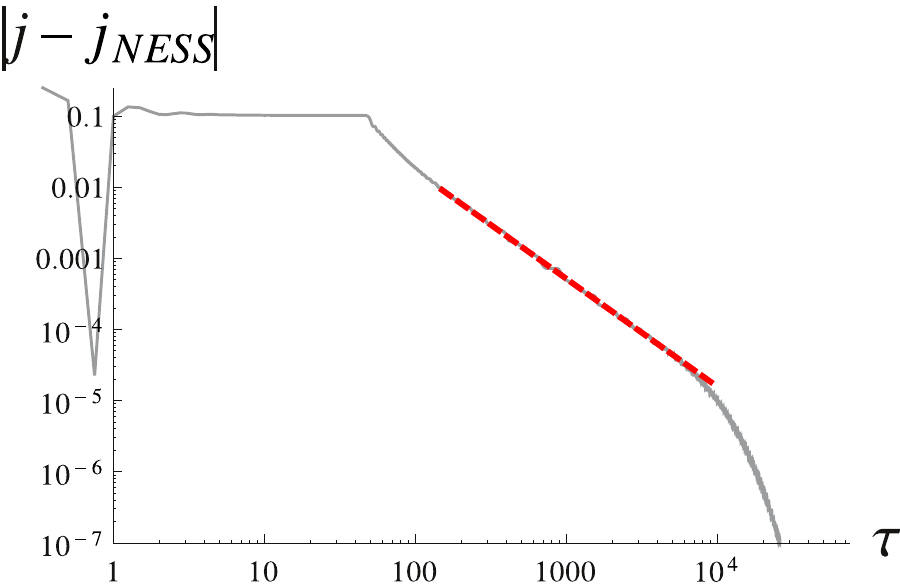}
B\includegraphics[width=0.85\linewidth]{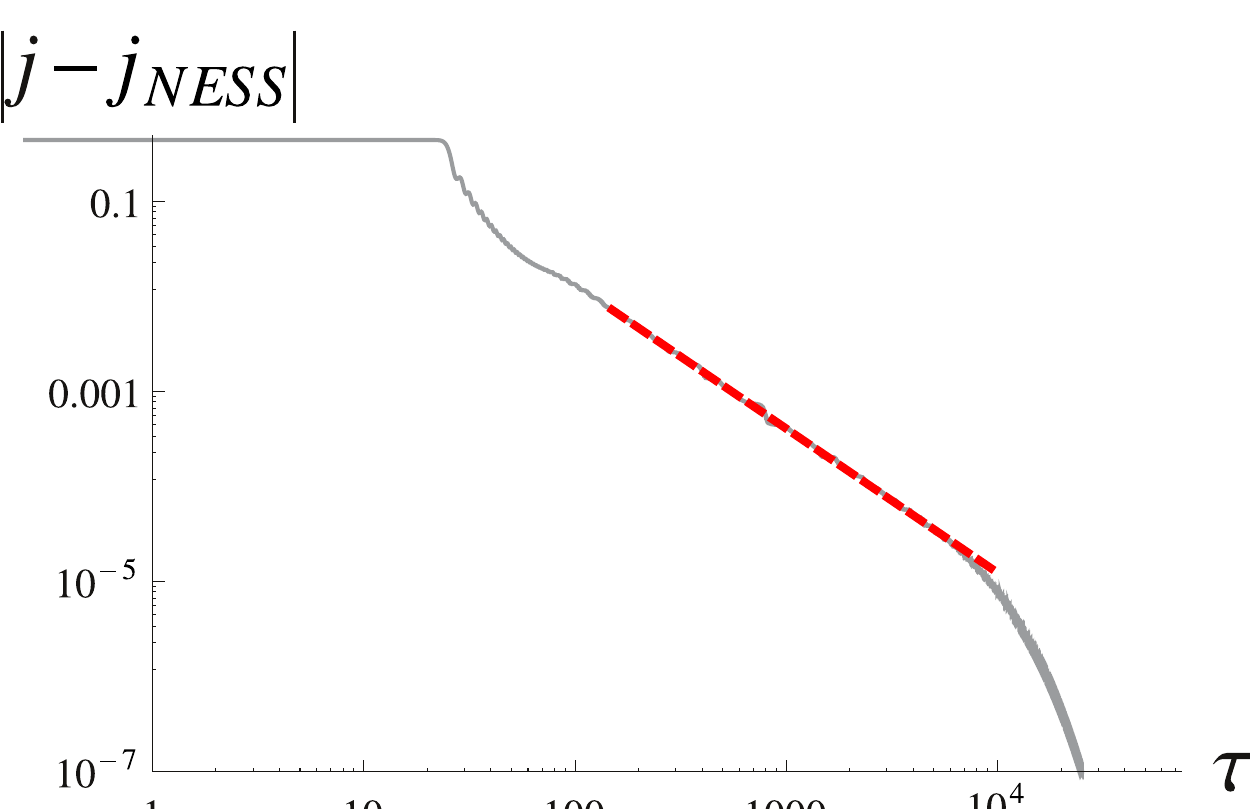}
C\includegraphics[width=0.85\linewidth]{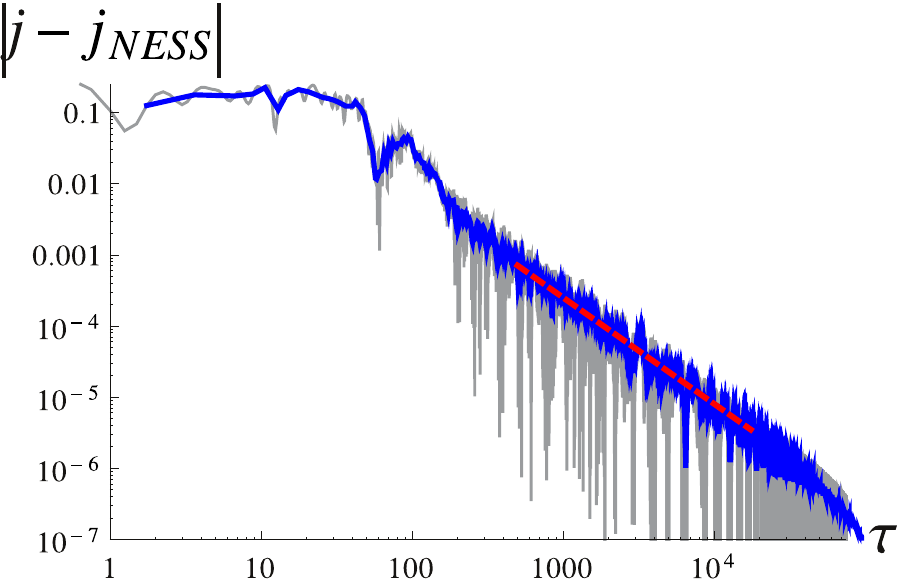}
\caption{\label{Eq_Current}
Dependence of the current on time in the open quantum system starting from A: completely empty chain, B: completely filled chain, C: some arbitrary density matrix with five diagonal and five off-diagonal elements. Gray line: raw data, blue line: data averaged over fast oscillations, red-dotted line: fit, 
 $|j-j_{NESS}| \propto\tau^{\nu}$. A: $\nu=-1.516\pm0.02$, B: $\nu=-1.515\pm0.02$, C: $\nu=-1.495\pm0.01$. 
}\end{center}
\end{figure}

The resulting equilibration can be separated into three stages: the initial current-building plateau, with the duration proportional to the length of the system, the intermediate fast decay, whose duration is also proportional to the  system length, and the long time power-law tail, which is followed by an exponential decay related to the finite length of the system. 
The power-law decay starts  when both the decay rates and the value of the current on the NESS-excitations can be approximated by the power-law behaviour: $\tau_{pow}\approx N\left(\frac{\Gamma_1^{(i)}}{1+\Gamma_1^{2(i)}}+\frac{\Gamma_N^{(o)}}{1+\Gamma_N^{2(o)}} \right)^{-1}$, $N\gg 1$.
The beginning of the exponential decay is estimated from the smallest imaginary part of $\lambda_j$, leading to the start of the exponential decay at $\tau_{exp}\approx{N^{3}}\left(\frac{\Gamma_1^{(i)}}{1+\Gamma_1^{2(i)}}+\frac{\Gamma_N^{(o)}}{1+\Gamma_N^{2(o)}} \right)^{-1}$, $N\gg 1$.  

\subsubsection{One-dimensional chain}

\begin{figure}[tb!]
\begin{center}
A\includegraphics[width=0.85\linewidth]{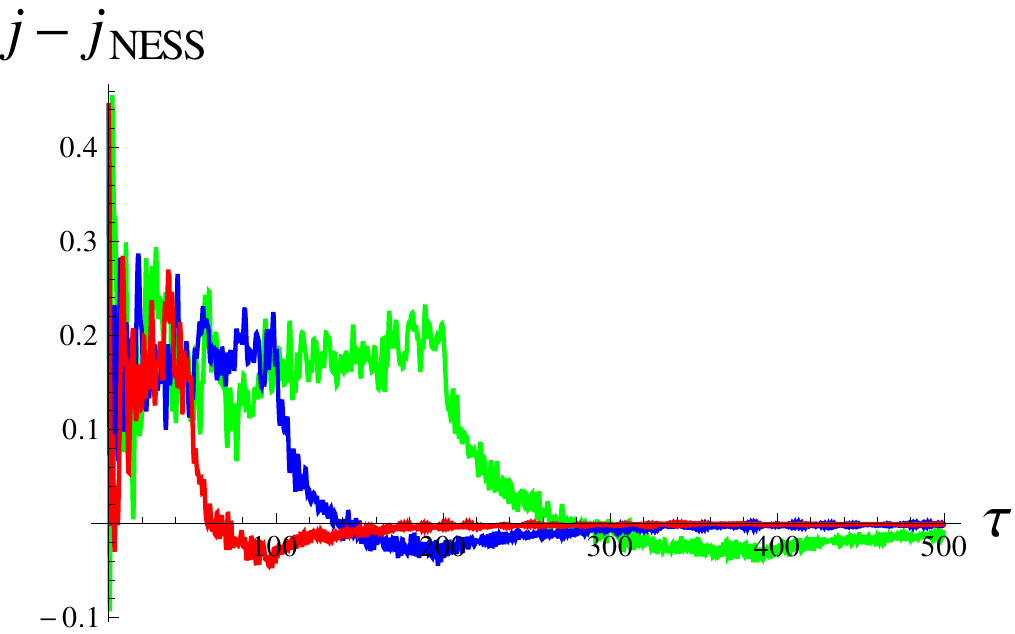}
B\includegraphics[width=0.85\linewidth]{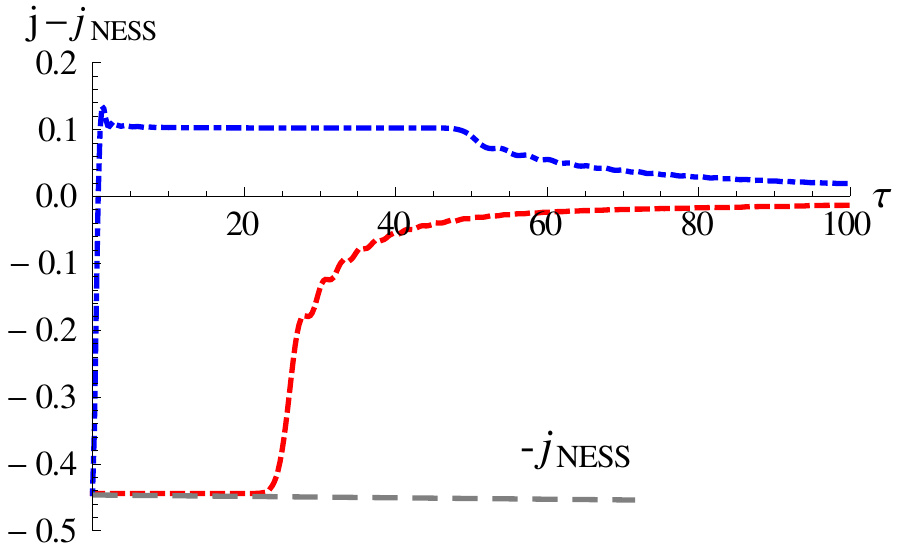}
C\includegraphics[width=0.85\linewidth]{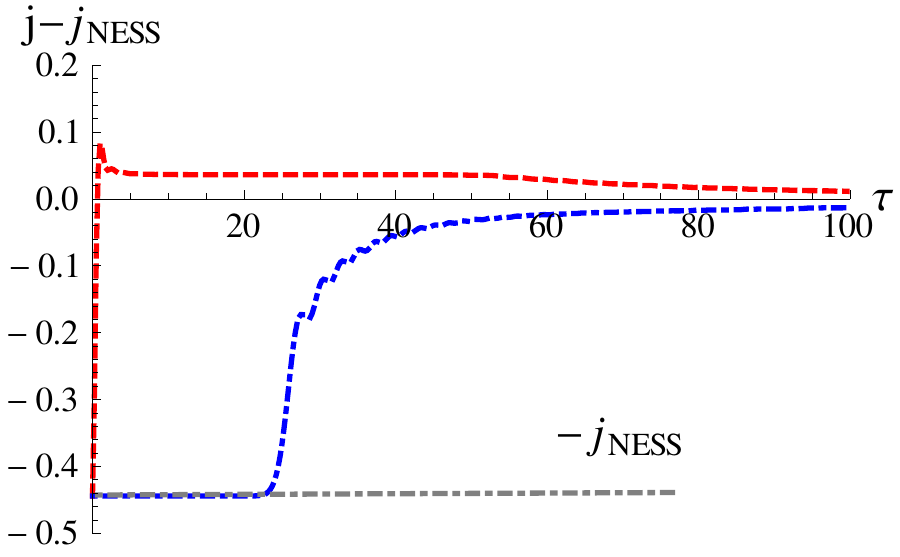}
\caption{\label{CurrentLength}
A: Equilibration of the current at early times for different lengths of the system $N=100,200,400$ with an arbitrary initial density matrix. The time of the initial equilibration is proportional to the length of the system.
B: The equilibration for the initially empty, C: for the initially filled. The blue dash-dotted line is a current through the first site of the chain, coupled to drain, the red dashed line represents the current through the last site of the chain, coupled to the source.
}\end{center}
\end{figure}

The initial equilibration plateau is seen clearly at Fig.~\ref{CurrentLength}A where the time dependence of the current is shown for different system lengths. 
Such behaviour is connected to the finite speed of propagation of excitations in the system. It is best illustrated  by analysing the time-dependence of the current through the first and the last site for the completely filled or completely empty chain, Fig.~\ref{CurrentLength}B,C. For a chain which is initially completely empty, we expect no current through the end of the chain coupled to the drain. The current through this end of the chain starts to flow only when the fastest excitation coming from the source reaches it. The dispersion relation of the excitations in zeroth order in $1/N$ is $\lambda(k)=2t\cos k$, where  $k=\tfrac{\pi j}{N+1}$, in accordance with Eqs.~(\ref{lambda0}), (\ref{deltaL2}). The phase velocity of the excitation is $v(k)=\tfrac{\pp \lambda}{\pp k}$. Therefore, the fastest excitation has the speed $2t$. (In the expressions of this section we have restored the energy scale $t$, which is the hopping matrix element of the tight-binding chain.) It travels through the whole system during the time $N/2t$, which is in correspondence with Fig.~\ref{CurrentLength}B, red dashed line.  Then the system starts to leak to the drain. This information again propagates to the opposite end (to the end connected to the source), end during the time $N/2t$. Hence, after the time $N/t$ the equilibration at the source end starts, Fig.~\ref{CurrentLength}B, blue dash-dotted line.
Therefore, the initial plateau is connected to the build-up of correlations in the system.

Similar considerations are applicable for the completely empty chain. In this case the current through the source end is initially zero, as there is no place in the chain for new particles. The first information about the emptying the chain through a connection to the drain reaches the source end in the time $N/2t$, the equilibration of the current through the first chain site starts at this time, Fig.~\ref{CurrentLength}C, blue dash-dotted line. The correlations through the system start to build up and reach the drain at time $N/t$. The equilibration towards the NESS value of the current also starts at this time, through the other (drain) end of the chain.   
Let us note that for the case of equal couplings to the source and the drain, the current through the first site of the empty chain is equal to the current through the last site of the full chain and vice versa.


\begin{figure}[tb!]
\begin{center}
\includegraphics[width=0.85\linewidth]{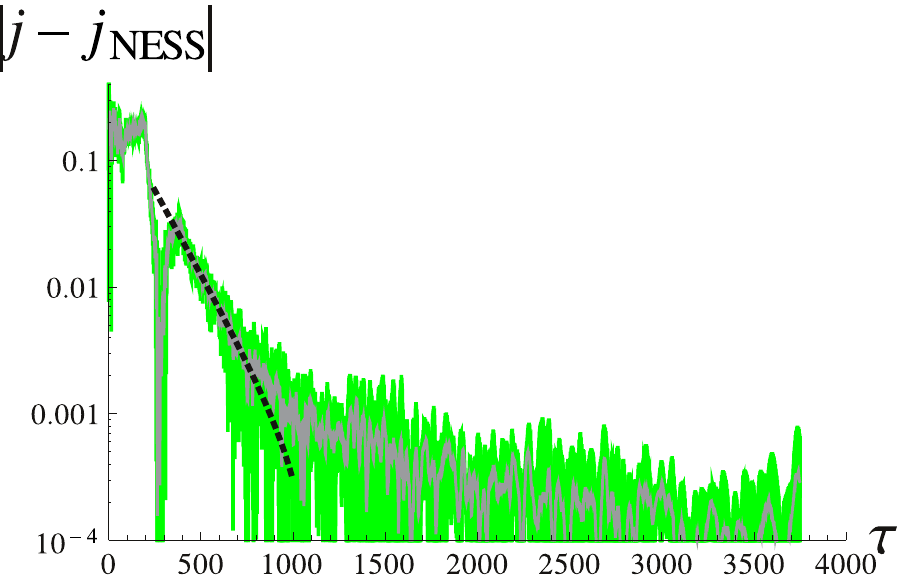}
\caption{\label{CurrentLength2}
Initial equilibration on the logarithmic scale shows the existence of the initial exponential decay region. Green: data for $N=400$, gray: averaged over fast oscillations, black dotted: exponential fit for time of order $N$.
}\end{center}
\end{figure}

The current-buildup plateau is generic for the initial states which did not have a uniform current passing through the system.   
The intermediate regime between the current-building plateau and the power-law relaxation regime is clearly seen in the log-linear plot, Fig.~\ref{CurrentLength2}. 
Its duration is determined by the $\delta\lambda_j$ which still do not follow a quadratic behaviour. 
Around time $\tau \sim \tau_{pow}$, the relaxation to the equilibrium slows down, Fig.~\ref{Eq_Current}. It is described by the power-law derived in the previous section, Eq.~(\ref{EqCurrent}). Indeed, the fit of the data averaged over the fast oscillations gives approximately $\tau^{-3/2}$ decay rate towards the NESS-value of the current. (The power-law exponent is determined from a numerical fit and depends slightly on the time interval in which we do the fit. We include this uncertainty in the errors bars as well.)

\begin{figure}[tb!]
\begin{center}
\includegraphics[width=0.85\linewidth]{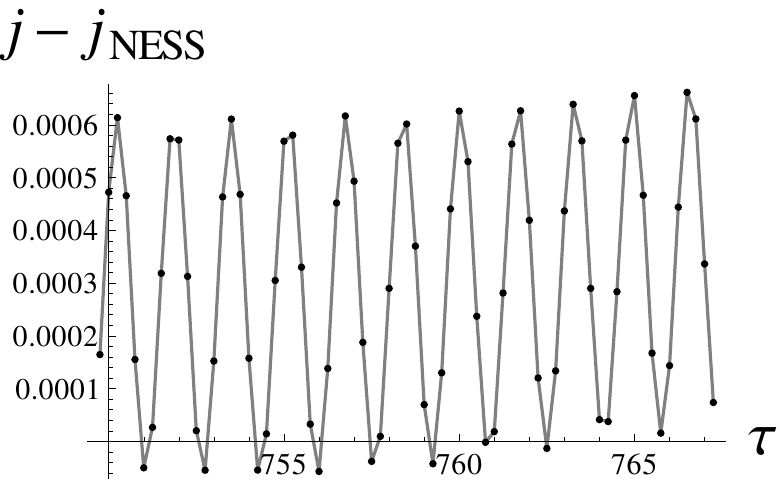}
\caption{\label{oscillations}
Zoom-in of the oscillations of the current. 
}\end{center}
\end{figure}

In the long-time limit only the modes with $k$ close $0$ and close $\pi$ survive, Fig.~\ref{dispersion}, Eq.~(\ref{deltaL2}). The NESS-excitations $\rho_{jm}^{(2)}$ in $\{f\}$-basis "connect" these two modes. They have an oscillating behaviour with the period  of order $2t$, Fig.~\ref{oscillations}.
The NESS is unique in our problem. The other name of NESS, as the state which does not evolve with time, is dark state. The NESS-excitations corresponding to the $k$ values close $0$ and close to $\pi$ can be named non-equilibrium dark states. In the steady state (NESS) they die out, but  while approaching NESS they live very long. When the system has several dark states, any linear combination of them is a dark state as well. For non-equilibrium dark states it is also true that their linear combination is a solution as well (though here it cannot be an arbitrary linear combination, as the overall density matrix should be positive definite, which limits the coefficients in the linear combination). 
But non-equilibrium dark states can as well be connected by the oscillating terms, which are the NESS-excitations $\rho_{jm}^{(2)}$.

\begin{figure}[tb!]
\begin{center}
\includegraphics[width=0.85\linewidth]{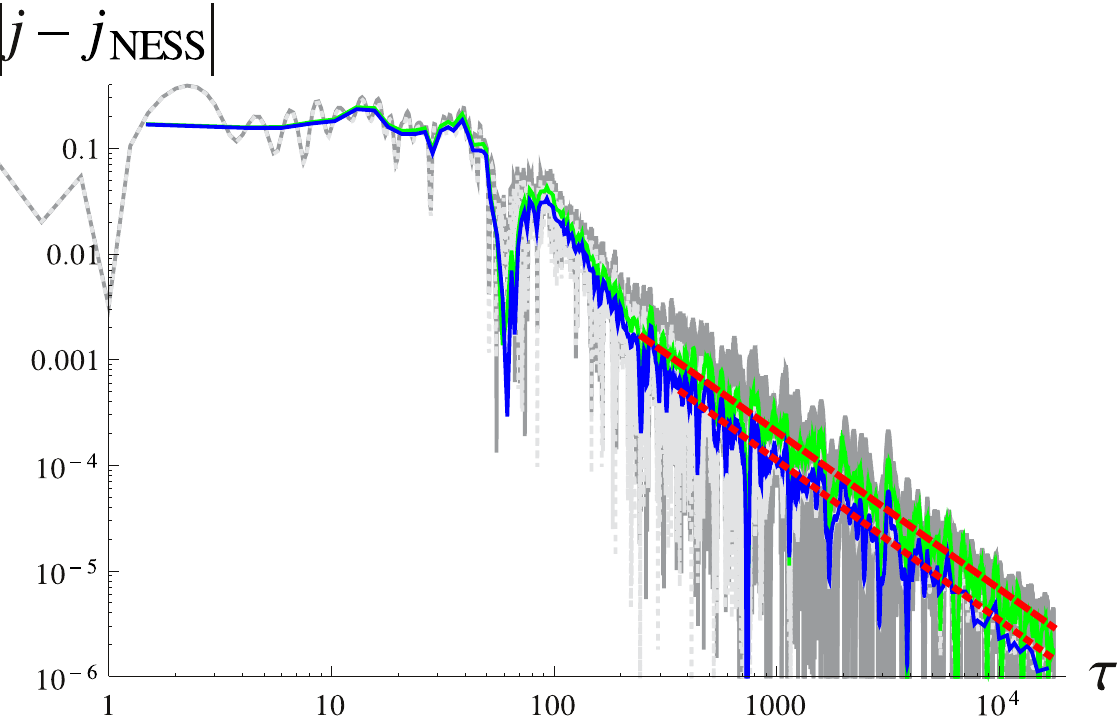}
\caption{\label{Eq_Decay_Current}
Difference $j(\tau)-j_{NESS}$ starting from some arbitrary density matrix with three diagonal and one non-diagonal element. Light gray line is the current taking into account the phenomenological decay rate of the high-energy modes, while dark gray is the evolution without it. Blue and green lines represent the current averaged over the fast oscillations without and with phenomenological decay rate. Dashed and dotted lines are the fits: $x^{-1.49\pm 0.02}$ and $x^{-1.51\pm 0.05}$.
}\end{center}
\end{figure}

One could argue that the modes with high momentum are not physically relevant in real-world systems. They can be coupled to some other  degrees of freedom, for example to phonons in condensed mater systems and dissipate via phonons, or just decay in optical lattices. 
This coupling would lead to damping. We can introduce the damping in the model "artificially", adding an imaginary part to $\lambda_i$, representing this additional damping for the modes with large momentum, 
Fig.~\ref{Eq_Decay_Current}. The oscillations in the long-time limit are decreased comparing to the case without oscillations. One expects it as the oscillation are connected to the density matrices $\rho_{jk}^{(2)}$ which become damped.

\subsubsection{Two-dimensional lattice}

\begin{figure}[tb!]
\begin{center}
A\includegraphics[width=0.85\linewidth]{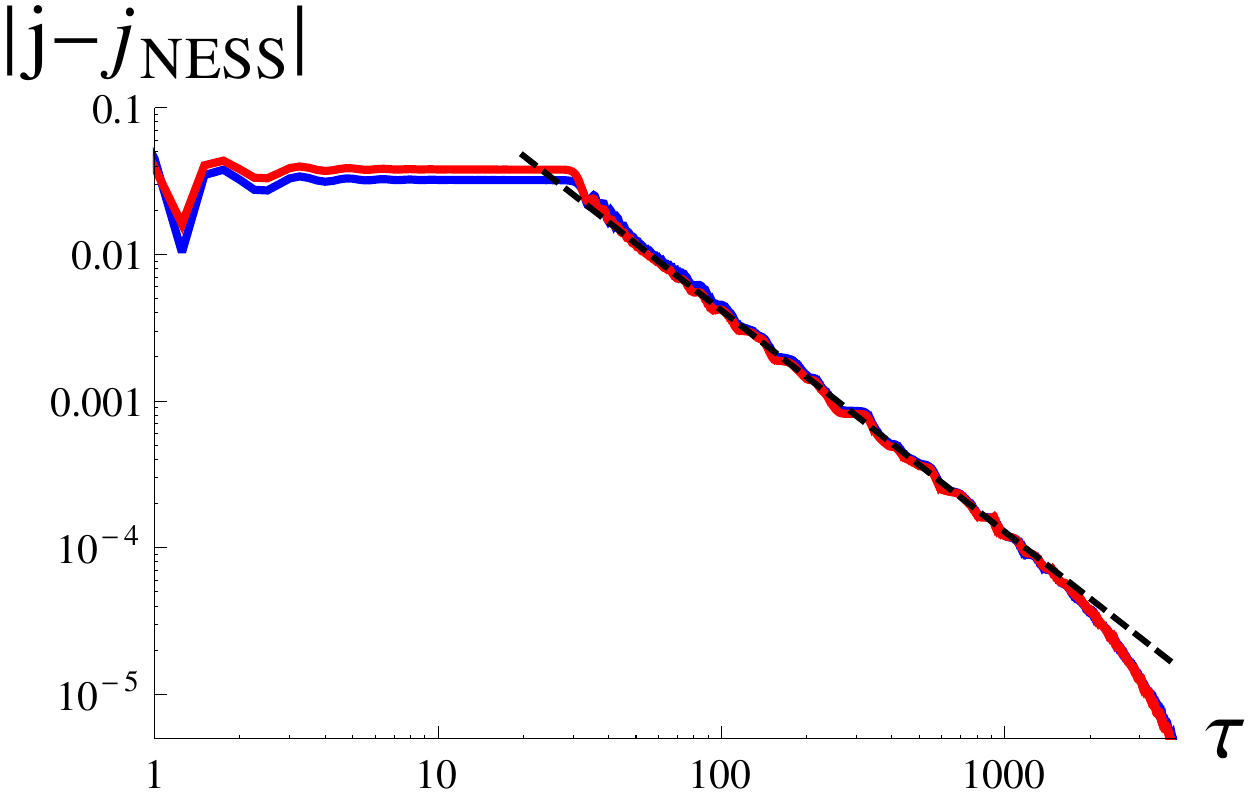}
B\includegraphics[width=0.85\linewidth]{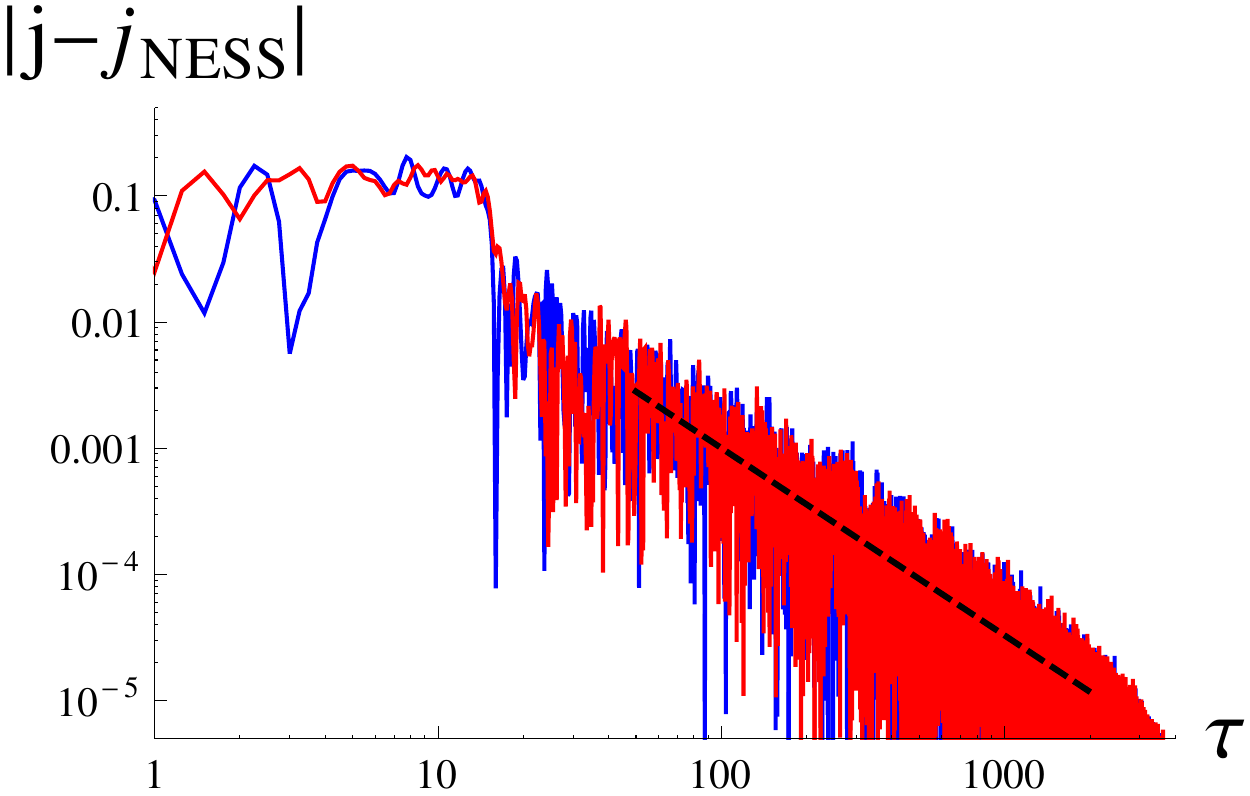}
\caption{\label{2d}
Current through a chain two-dimensional lattice of the size $N$ by $N$ coupled at two ends to the memoryless reservoirs with $\G_{1st~raw}^{(i)}=0.3$ and $\G_{Nth~raw}^{(o)}=0.6$, $N=29$. The current at the edges of the strip is almost the same as in the middle (blue and red lines). 
A: initially empty chain, B: an initial density matrix has three arbitrary non-zero diagonal elements. The fit gives exponent of the power-law decay A: $x^{1.51\pm0.02}$, B: $x^{1.49\pm0.05}$ .
}\end{center}
\end{figure}

Here we give a relaxation of the current through a two-dimensional strip: we take a rectangular strip coupled where each site at two opposite sides is coupled to the source and to the drain with the same coupling strengths. The simulation is done in a similar way as for the one-dimensional system, Fig.~\ref{2d}. We obtain the same power-law decay towards the equilibrium as in one-dimensional case.  

We have expected the same power-law in the current relaxation also from analytical considerations. The conclusion is that the analytical considerations used in the previous section are valid for non-interacting fermions in any dimension and lead to a independent of dimension  power-law relaxation rate in long-time dynamics of equilibration. 

\section{Conclusions}

 In this paper we consider the equilibration in open quantum systems by representing a time-dependent density matrix as an expansion around the NESS density matrix. This expansion consists of  NESS-excitations, each of which has a fixed decay rate. In closed quantum systems, each excitation can be a state of the system. NESS-excitations themselves cannot represent a state of the system, as they are traceless matrices. They are meaningful only when they are added to the NESS-density matrix, which by definition has unit trace.  

The expansion in NESS-excitations gives us a general structure of the time-dependent density matrix. This provides a way of understanding the time-dependence of observables. The decay of any state to the NESS is given by the sum over the values of the observables on the NESS-excitations and the density of states of the decay rates of the NESS-excitations. 

We illustrate our viewpoint on equilibration on the example of a chain of non-interacting fermions linearly coupled to memoryless reservoirs. For this system the calculation of the NESS-excitations and their decay rates can be performed analytically. 
We identify three time scales in the time evolution of the current: i) plateau of buildup for times proportional to the length of the system; ii) faster-than-power-law decay of the modes in the middle of the spectrum; iii) algebraic decay towards the NESS-value of the current, which turns into the exponential
at times proportional to the third power of the system length for
non-interacting systems due to the finite system size.
The exponent of the power-law decay for non-interacting fermions in the rectangular geometry does not depend on the dimension and has $\tau^{-3/2}$-dependence.

The observables also experience universal oscillations with the period $\tfrac{\pi}{t m}$ super-imposed on the power-law decay due to the contribution of NESS-excitations which have an oscillating behaviour, as for example $\rho_{1/2,jk}^{(2)}$, and where $m$ is the  order of the contributing NESS-excitation.

In the case of interacting fermions the exact form of NESS-excitations is more complicated as it is impossible to bring into the Liouvillian into the diagonal form. The one-particle decay rate in the $\{f\}$-basis will be modified by both the coupling to the reservoirs and the interaction between the particles. At low interaction strength the system has effective quasiparticles. Then the NESS-excitations can be constructed starting from the effective quasiparticles. The power-law decay towards the NESS is still present, but the time-scale when it finishes is limited by the interaction strength.

An interesting question which we have not touched upon here is the influence of the geometry on the current relaxation. For example, would the exponent of the power-law relaxation change for the non-rectangular strips?
For example, a current through a point junction connecting two non-interacting fermionic reservoirs has been studied recently~\cite{Mag} and the power law which determines the relaxation is $\tau^{-2}$ at high temperatures and $\tau^{-1}$ at low temperatures in the reservoirs which is different from our $\tau^{-3/2}$ approach to the NESS.

The power-law equilibration can be observed in the transport experiments in cold-atom systems which have recently been carried out.\cite{OptLattice} The difference with respect to existing setups is that for testing of our-predictions one should make a lattice of fermions in the transport region.

\section*{Acknowledgments}
This work was supported through SFB 1073 of the Deutsche Forschungsgemeinschaft (DFG).
We are grateful for reading the manuscript to Mihailo \u{C}ubrovi\'{c}, Fabian Biebl, Salvatore Manmana. 

\appendix
\begin{widetext}
\section{Transformation of the Liouvillian to the diagonal basis}
\label{sec:LTransformation}

We consider a lattice of non-interacting fermions described by the tight-binding Hamiltonian:
\be H=  \sum_{\la i,j \ra} t \left( a^\dg_i a_{i+1}+ h.c \right)+ \sum_i U_i a^\dg_i a_i , \nonumber\ee
where $t$ is the hopping matrix element between the neighbouring sites,  $U_i$ is an on-site potential and $\la i,j \ra$ stands for the connectivity of the lattice.  
Evolution of a system coupled to the memoryless bath is described by the Lindblad equation:
\be i\frac{d\rho}{d\tau}= [H,\rho] + 
i \sum_{\mu,i/o}\left(2 L^{(i/o)}_\mu  \rho L^{\dg,(i/o)}_\mu - \{L^{(i/o)}_\mu L^{\dg,(i/o)}_\mu,\rho\} \right)\nonumber\ee
where $L_\mu$ are the Lindblad operators responsible for the coupling to the bath. The lattice of fermions is coupled to the source and the drain at infinite bias voltage~\cite{Gurv1,Gurv2} at the ends of the chain: $L_{\{in\}}^{(i)}=\sqrt{\Gamma_1^{(i)}}a_{\{in\}}^\dg$, $L^{(o)}_{\{out\}}=\sqrt{\Gamma_N^{(o)}}a_{\{in\}}$.  

The solution of the Lindblad equations for non-interacting fermions is notably simplified in the super-fermionic representation~\cite{Kosov, Medv}: operators acting from the right on the density matrix are introduced. They are denoted by "tilde". Then the Liouvillian can be written after the particle-hole transformation
$ \td{a}=b^\dg, \td{a}^\dg=b$ in the quadratic form:
\be \label{Ldiagonal}
\mathcal{L} = (a^\dg b^\dg) \mathcal{M} \begin{pmatrix}
                               a \\
                               b
                              \end{pmatrix}
 -i\sum_\mu \Gamma_\mu^{out} - i\sum_\mu \Gamma_\mu^{in}.\ee
For the one-dimensional chain with $L_1^{(i)}=\sqrt{\Gamma_1^{(i)}}a_1^\dg$, $L^{(o)}_N=\sqrt{\Gamma_N^{(o)}}a_N$ the matrix $\mathcal{M}$ has a structure:
\be 
\label{m} \mathcal{M} = \begin{pmatrix}[ccccc|ccccc]
                        -i\Gamma^{(in)}_1+U & t & 0 & \ldots& 0 &0 &\ldots&\ldots&\ldots& 0 \cr
                        t & U & t & \ddots & \vdots & 0  & \ldots&\ldots&\ldots& 0 \cr
                        0 & \ddots & \ddots & \ddots & \vdots & 0& \ldots& \ldots&\ldots& 0 \cr                
                        \vdots &  \ldots &\ddots & U & t & 0 & \ldots & \ldots&\ldots& 0\cr
                        
                        0 & \ldots & 0 & t & i\Gamma^{(out)}_2+U & 0 & \ldots & \ldots& 0& 2\Gamma^{(out)}_2 \cr
                        \hline
                        -2\Gamma^{(in)}_1 &0 &\ldots&\ldots& 0& i\Gamma^{(in)}_1+U & t & 0 & \ldots& 0  \cr
                        0 &\ldots&\ldots&\ldots& 0& t & U & t & \ddots & \vdots  \cr
                        0 &\ldots&\ldots&\ldots& 0 & 0 & \ddots & \ddots & \ddots & \vdots \cr                
                        0 &\ldots&\ldots&\ldots& 0 & \vdots & \ldots & \ddots & U & t\cr
                        
                         0 &\ldots&\ldots&\ldots& 0& 0 & \ldots & 0 & t & -i\Gamma^{(out)}_2+U  
                       \end{pmatrix}
 \ee

Due to this specific structure of $\mathcal{M}$ the constant terms in the expression~(\ref{Ldiagonal})  vanish after introducing a new set of the operators  $\{f,f^\dgc,\td{f},\td{f}^\dgc\}$~\cite{Medv} and even more in this basis the Liouvillian becomes diagonal:
\be \mathcal{L}_f = \sum_i \lambda_i f_i^\dgc f_i - \sum_i \lambda_i^* \td{f}_i^\dgc \td{f}_i. \nonumber \ee
The operators $\{f^\dgc,\td{f}^\dgc\}$ are dual to the operators $\{f,\td{f}\}$, but not 
Hermitian conjugated, though the operators obey anti-commutation relations. 
The operators $\{f,f^\dgc,\td{f},\td{f}^\dgc\}$ are linear combinations of the operators $\{a,a^\dg,\td{a},\td{a}^\dg\}$:
\bea
a_m^\dg = \sum_{k_1}C^{(1)}_{mk_1}f^\dgc_{k1} + C^{(2)}_{mk_1}\td{f}_{k1}, \nonumber \\
a_m = \sum_{k_1}A^{(1)}_{mk_1}f_{k1} + A^{(2)}_{mk_1}\td{f}^\dgc_{k1}. \nonumber
\eea 
The coefficient matrices $C$ and $A$ are connected to the matrix of the eigenvectors $P$ of the matrix $\mathcal{M}$ (See Ref.~\onlinecite{Medv}):
\be P = \begin{pmatrix}[cc]
A^{(1)} & A^{(2)} \cr
A^{(3)} & A^{4)}
\end{pmatrix}, 
(P^{-1})^T = \begin{pmatrix}[cc]
C^{(1)} & C^{(2)} \cr
C^{(3)} & C^{(4)}
\end{pmatrix}.\ee
In $P$ the eigenvectors are ordered in the following way: first $N$ of eigenvectors correspond to eigenvalues with a negative imaginary part, while the the second half have a positive imaginary part and are complex conjugated to the first set. All matrices  $A^{(i)}$ and $C^{(i)}$, $i=1,\ldots,4$ have dimension $N\times N$.

In this formalism, instead of solving a differential equation for the evolution of the $2^N\times 2^N$  density matrix, the calculations are done with the $2N\times 2N$ matrices. 

The values $\lambda_i$ are the eigenvalues of the matrix $\mathcal{N}$~\cite{Medv} (we put $U_i=0$):
\be  \label{N} \mathcal{N} = \begin{pmatrix}[ccccc]
                        -i\Gamma_1^{(i)} & t & 0 & \ldots& 0  \cr
                        t & 0 & t & \ddots & \vdots  \cr
                        0 & \ddots & \ddots & \ddots & \vdots  \cr                
                        \vdots &  \ldots &\ddots & 0 & t \cr
                       
                        0 & \ldots & 0 & t & -i\Gamma_N^{(o)}
                       \end{pmatrix}.
\ee

The density matrix of NESS is the vacuum for the operators $f$ and $\td{f}$: $\rho_{NESS}=|00\ra_{f,\td{f}}$. Therefore, the observables in NESS can be computed directly by transforming an operator from the $\{a\}$-basis to the  $\{f\}$-basis.

\section{NESS-excitations for the two-site problem}
\label{sec:example}
  
Here we write down explicitly the NESS-excitations and their decay rates first in terms of the creation  operators in the  $\{f\}$-basis, Table~\ref{tab:twosite}, and second we represent the NESS-excitations in $\{a\}$-basis, Eqs.~(\ref{EqBeg})-(\ref{EqEnd}). It can be done in two ways. 
The first way involves as the first step the calculation of the NESS density matrix in the $\{a\}$-basis and then acting on it with creation operators in the $\{f\}$-basis which are represented as operators in the $\{a\}$-basis. The second way is to work explicitly with the Liouvillian of the problem which is a 16 by 16 matrix, finding its eigenvalues and constructing the combinations from its eigenvectors which would have the same decay rate and obey properties of NESS-excitations.  In this Appendix we measure energy in the units of $t$, thus $t=1$.

\begin {table}[htb]
\normalsize
\caption {NESS-excitations in $\{f\}$-basis and their decay rates for a two-site chain.} 
\label{tab:twosite} 
\begin{center}
\begin{tabular}{|c|c|c|c|}
\hline
$N$ & decay rate & NESS-excitation $\hat{F}$ in the $\{f\}$-basis & Notation \\
 &  & $\hat{F} |00\rangle_{f,\td{f}}$ & for $\rho$ \\
\hline
0 & 0 & $\hat{1}$ & $\rho_0$
\\ \hline
1& $-\tfrac{1}{2}(\Gamma_1+\Gamma_2) \pm \tfrac{i}{2}\sqrt{ - (\Gamma_1-\Gamma_2)^2} \pm i U$ & 
$ i (e^{i\omega_1 \tau}f_1^\dgc + e^{-i\omega_1 \tau} \gamma \td{f_1}^\dgc)$, $\omega_1= \delta/2 + U$  & 
$\rho_{1,1}$ \\ 
1& $-\tfrac{1}{2}(\Gamma_1+\Gamma_2) \mp \tfrac{i}{2}\sqrt{4 - (\Gamma_1-\Gamma_2)^2} \pm i U$ & 
$i(e^{i\omega_2 \tau} f_2^\dgc  + e^{-i\omega_2 \tau} \gamma \td{f_2}^\dgc )$, $\omega_2= -\delta/2+ U$  & 
 $\rho_{1,2}$ \\ \hline
2& $-(\Gamma_1+\Gamma_2)$ twice & $e^{i \phi} f_1^\dgc \td{f_1}^\dgc, e^{-i\phi}f_2^\dgc \td{f_2}^\dgc, \phi=\arctan \tfrac{\Gamma_1+\Gamma_2}{\delta}$ & $\rho_{2,1}$, $\rho_{2,2}$\\ \hline
2& $-(\Gamma_1+\Gamma_2) \pm 2 i U$ & $e^{i\omega_3 \tau}f_1^\dgc f_2^\dgc  + e^{-i\omega_3 \tau} |\gamma|^2 \td{f}_1^\dgc \td{f}_2^\dgc $,  $\omega_3=2U$&$\rho_{2,3}$\\ \hline 
2& $-(\Gamma_1+\Gamma_2) \pm i\sqrt{4 - (\Gamma_1-\Gamma_2)^2}$ & $e^{i\omega_4 \tau}f_1^\dgc \td{f_2}^\dgc + e^{-i\omega_4 \tau}f_2^\dgc \td{f_1}^\dgc$ , $\omega_4=\delta$ & $\rho_{2,4}$ \\ \hline
3&  $-\tfrac{3}{2}(\Gamma_1+\Gamma_2) \pm \tfrac{i}{2}\sqrt{4 - (\Gamma_1-\Gamma_2)^2} \pm i U$ & $f_2^\dgc \td{f_2}^\dgc(e^{i\omega_1 \tau}f_1^\dg +e^{-i\omega_1 \tau}\gamma\td{f_1}^\dgc)$  & $\rho_{3,1}$\\ 
3&  $-\tfrac{3}{2}(\Gamma_1+\Gamma_2) \mp \tfrac{i}{2}\sqrt{4 - (\Gamma_1-\Gamma_2)^2} \pm i U$ & $f_1^\dgc \td{f_1}^\dgc( e^{i\omega_2 \tau}f_2^\dg + e^{-i\omega_2 \tau}\gamma\td{f_2}^\dgc)$  & $\rho_{3,2}$\\ \hline
4& $-2(\Gamma_1+\Gamma_2)$ &  $f_1^\dg \td{f_1}^\dgc f_2^\dg \td{f_2}^\dgc$  & $\rho_4$\\
\hline
\end{tabular}
\end{center}
\end{table}

\be \label{EqBeg}
\rho_0=\frac{\Gamma_1 \Gamma_2}{(\Gamma_1+\Gamma_2) (1+\Gamma_1 \Gamma_2)}\begin{pmatrix}
        \frac{\Gamma_2}{\Gamma_1(\Gamma_1+\Gamma_2)} & 0 & 0 & 0 \\\
        0 & \frac{\left(1+(\Gamma_1+\Gamma_2)^2\right)}{(\Gamma_1+\Gamma_2)} & i & 0 \\
         0 & -i & \frac{1}{(\Gamma_1+\Gamma_2)} & 0 \\
         0 & 0 & 0 & \frac{\Gamma_1}{(\Gamma_1+\Gamma_2)\Gamma_2}
      \end{pmatrix} 
\ee      

\bea
\rho_{1,1}(\tau)&=& \exp(-(\Gamma_1+\Gamma_2)\tau/2)\left( e^{-i\omega_1 \tau}R_{1,1}+e^{i\omega_1 \tau}R^\dg_{1,1} \right)\\
&&R_{1,1}=
                                                  \begin{pmatrix} 
                                                  0 & 0 & 0 & 0 \\
                                                  X_1 & 0 & 0 & 0 \\
                                                  X_2 & 0 & 0 &  0\\
                                                  0& X_3 & X_4 & 0
                                                 \end{pmatrix}, \nonumber\\
&& X_1=\Gamma_2 (-2-\Gamma_1^2+\Gamma_2^2)-i\delta \Gamma_2(\Gamma_1+\Gamma_2), 
 X_2=-\delta\Gamma_2+i\Gamma_2(\Gamma_1-\Gamma_2),\nonumber\\
&& X_3=\Gamma_1(-\delta-i(\Gamma_1+3\Gamma_2)),
 X_4=2\Gamma_1.\nonumber
 \eea

\bea
\rho_{1,2}(\tau)&=& \exp(-(\Gamma_1+\Gamma_2)\tau/2)\left( e^{-i\omega_2 \tau}R_{1,2}+e^{i\omega_2 \tau}R^\dg_{1,2} \right)\\
&&R_{1,2}=
                                                  \begin{pmatrix} 
                                                  0 & 0 & 0 & 0 \\
                                                  X_1^{*} & 0 & 0 & 0 \\
                                                  -X_2^{*} & 0 & 0 & 0 \\
                                                  0& -X_3^{*} & X_4 & 0
                                                 \end{pmatrix}.\nonumber
 \eea

\bea
\rho_{2,1}(\tau)&=& \exp(-(\Gamma_1+\Gamma_2)\tau)\begin{pmatrix}
        1 & 0 &0 &0\\
        0& \frac{1}{2}\left(-1+\frac{\Gamma_1}{\Gamma_2}\right) & \frac{i(\Gamma_1^2-\Gamma_2^2)-\delta(\Gamma_1+\Gamma_2)}{4\Gamma_2} &0 \\
        0 & \frac{-i(\Gamma_1^2-\Gamma_2^2)-\delta(\Gamma_1+\Gamma_2)}{4\Gamma_2}&\frac{1}{2}\left(-1+\frac{\Gamma_1}{\Gamma_2}\right)& 0 \\
        0 & 0& 0 & -\frac{\Gamma_1}{\Gamma_2} 
       \end{pmatrix}\\    
\rho_{2,2}(\tau)&=& \exp(-(\Gamma_1+\Gamma_2)\tau)\begin{pmatrix}
       1 & 0 &0 &0\\
        0& \frac{1}{2}\left(-1+\frac{\Gamma_1}{\Gamma_2}\right) & \frac{i(\Gamma_1^2-\Gamma_2^2)+\delta(\Gamma_1+\Gamma_2)}{4\Gamma_2} &0 \\
        0 & \frac{-i(\Gamma_1^2-\Gamma_2^2)+\delta(\Gamma_1+\Gamma_2)}{4\Gamma_2}&\frac{1}{2}\left(-1+\frac{\Gamma_1}{\Gamma_2}\right)& 0 \\
        0 & 0& 0 & -\frac{\Gamma_1}{\Gamma_2} 
      \end{pmatrix}
\eea

\bea     
\rho_{2,3}(\tau)&= &\exp(-(\Gamma_1+\Gamma_2)\tau)\left(e^{i\omega_3 \tau}\begin{pmatrix}
       0 & 0 &0 & 0\\
       0 & 0 & 0 &0 \\
       0 &0 &0 & 0\\
       1& 0 &0 &0 
      \end{pmatrix}+ e^{-i\omega_3 \tau}\begin{pmatrix}
       0 & 0 &0 & 1\\
       0 & 0 & 0 &0 \\
       0 &0 &0 & 0\\
       0& 0 &0 &0 
      \end{pmatrix}\right)
\eea

\bea      
\rho_{2,4}(\tau)&= &\exp(-(\Gamma_1+\Gamma_2)\tau)\left(e^{i\omega_4 \tau}\begin{pmatrix}
       Y_1 & 0 &0 & 0\\
       0 & Y_2 & Y_5 &0 \\
       0 & Y_5^* & Y_3 & 0\\
       1& 0 &0 &Y_4 
      \end{pmatrix}+ e^{-i\omega_4 \tau}\begin{pmatrix}
       Y_1^* & 0 &0 & 1\\
       0 & Y_2^* & -Y_5^* &0 \\
       0 &- Y_5 &Y_3^* & 0\\
       0& 0 &0 & Y_4^* 
      \end{pmatrix}\right)\\
      &&Y_1= -\G_2(-2+(\G_1-\G_2)(\G_1-\G_2-i\delta)) , \\
&& Y_2 = -\G_1 - 3\G_2 + \G_1^2\G_2 -2\G_1\G_2^2 + \G_2^3 - i\delta(1+\G_2(\G_1 +\G_2)),\\
&& Y_3 = -\G_1 + \G_2 + i\delta,~~~ Y_4=2\G_1, ~~~ Y_5=\delta \G_2 - i (2+\G_1 \G_2 -\G_2^2).
\eea

\bea
\rho_{3,1}(\tau)&=& \exp(-3(\Gamma_1+\Gamma_2)\tau/2)\left( e^{i\omega_1 \tau}R_{3,1}+e^{-i\omega_1 \tau}R^\dg_{3,1} \right)\\
&&R_{3,1}=
                                                  \begin{pmatrix} 
                                                  0 & 1 & \tfrac{\delta}{2}+\tfrac{i}{2}(\G_1-\G_2) & 0 \\
                                                  0 & 0 & 0 & -\tfrac{\delta}{2}-\tfrac{i}{2}(\G_1-\G_2) \\
                                                  0 & 0 & 0 & 1 \\
                                                  0& 0 & 0 & 0
                                                 \end{pmatrix}.\nonumber
 \eea

\bea
\rho_{3,2}(\tau)&=& \exp(-3(\Gamma_1+\Gamma_2)\tau/2)\left( e^{i\omega_2 \tau}R_{3,1}+e^{-i\omega_2 \tau}R^\dg_{3,1} \right)\\
&&R_{3,1}=
                                                  \begin{pmatrix} 
                                                  0 & 1 & -\tfrac{\delta}{2}+\tfrac{i}{2}(\G_1-\G_2) & 0 \\
                                                  0 & 0 & 0 & \tfrac{\delta}{2}-\tfrac{i}{2}(\G_1-\G_2) \\
                                                  0 & 0 & 0 & 1 \\
                                                  0& 0 & 0 & 0
                                                 \end{pmatrix}.\nonumber
 \eea

\bea      
\label{EqEnd}
\rho_4(\tau)&=&\exp(-2(\Gamma_1+\Gamma_2)\tau)\begin{pmatrix}
    1 & 0 &0 & 0\\
       0 & -1 & 0 &0 \\
       0 &0 &-1 & 0\\
       0& 0 &0 &1
    
      \end{pmatrix}
\eea

\section{Hierarchy of excited density matrices}
\label{sec:hierarchy}

Let us discuss the order of the excited density matrices. We have considered the example of the two-site system in Appendix~\ref{sec:example}.
What is the order of different density matrix excited states for a long chain system?

For the setup we consider -- a tight-binding chain coupled to the reservoirs at infinite bias voltages -- the frequencies $\lambda_j$ in the diagonal form of the Liouvillian~(\ref{LDiagonal}) come in pairs $\lambda=\pm a+i b$.  

Let us assume the following order of the eigenvalues: 
\be \im \lambda_{m_1}=\im \lambda_{m_2}< \im \lambda_{m_3}=\im \lambda_{m_4}< \im \lambda_{m_5} <\ldots <\im\lambda_{m_N}.\ee
There are two NESS-excitations which correspond to the smallest decay rate $\im \lambda_{m_1}$:\be \alpha_1 f_{m_1}^\dgc + \beta_1 \td{f}_{m_1}^\dgc|00\ra_f,
\alpha_2 f_{m_2}^\dgc + \beta_2 \td{f}_{m_2}^\dgc|00\ra_f, \label{lowest}\ee
where the coefficients $\alpha_i$ and $\beta_i$ restore the hermiticity in the $\{a\}$-basis. 

The next NESS-excitation corresponds to either the quadratic expression in $f_{m_1}$ and $f_{m_2}$ (see the NESS-excitations $2$ from the table in Appendix~\ref{sec:example}) if $2\im \lambda_{m_1}< \im \lambda_{m_3}$,   or to the NESS-excitations of the type~(\ref{lowest}) with the operators $f_{m_3}$ and $f_{m_4}$.

For one-dimensional system $3\im \lambda_{m_1} < \im \lambda_{m_3}$ 
(in the thermodynamic limit $N\rightarrow \infty$ there is even a degeneracy $4\im \lambda_{m_1}  \le \im \lambda_{m_3}$, see Eq.~\ref{deltaL2_2}). Therefore, the first, second and third states correspond to the NESS-excitations $1,~2,~3$ of the table in Appendix ~\ref{sec:example}.
The next NESS-excitation is of the type~(\ref{lowest}) with the operators with numbers $m_3$ and $m_4$. Then follow the states  
$$e^{i\omega' \tau}f_{m_{1/2}}^\dgc f_{m_{3/4}}^\dgc  + e^{-i\omega' \tau} |\gamma|^2 \td{f}_{m_{1/2}}^\dgc \td{f}_{m_{3/4}}^\dgc ,  
e^{i\omega'' \tau}f_{m_{1/2}}^\dgc \td{f}_{m_{3/4}}^\dgc + e^{-i\omega'' \tau}f_{m_{1/2}}^\dgc \td{f}_{m_{3/4}}^\dgc.$$
Afterwards come the states of type $3$ from the table in Appendix~\ref{sec:example} with the index $m_1$ or $m_2$ for the operators which enter twice with the same index and with index $m_3$ or $m_4$ for the other. All excited density matrices can be constructed by similar induction arguments. 

The spectrum of the initial Liouvillian~(\ref{LindGen}) consists of the combinations of sets of different $\lambda_j$. 
The states corresponding to $\lambda_j$ itself have the form~(\ref{lowest}), the overall number of these states is $N$.  The states corresponding to the sets of two $\lambda_k$ have the structure of the states $2$ from the table in Appendix~\ref{sec:example}, there are $N^2$ of those.  For the sets consisting of three $\lambda$ values there are $C_N^3$ variants with all three $\lambda$ values different, $3C_N^3$ variants with all three $\lambda$ different and one of the operators taken as tilde, and  $2C_N^2$ variants with only two different $\lambda$ values, which gives in total $(N-1) N (2N-1)/3$ excited density matrices. The further density matrices can be constructed in a similar combinatorial way. 



\section{Eigenvalues and eigenvectors of the matrix $\mathcal{M}$}

\subsection{Eigenvalues}
\label{sec:EigenValues}

The eigenvalues of the matrix $\mathcal{N}$ can be computed perturbatively in the thermodynamic limit, $N\rightarrow \infty$. Indeed, for $\Gamma_1^{(i)}=0$ and $\Gamma_N^{(o)}=0$, the eigenvalues and the eigenvectors of the matrix $\mathcal{N}$ are known:
\be \lambda^{(0)}_j= 2\cos \frac{\pi j}{N+1}, \phi^{(0)}_{j,k} = \sqrt{\frac{2}{N+1}}\sin \frac{\pi j k}{N+1}, \label{corrections} \ee 
where each of the eigenvectors $\phi_j$ is normalized to one. (In this Appendix the energy unit is $t$, thus $t=1$.)
The first-order in $\Gamma_1^{(i)}$ and $\Gamma_N^{(o)}$ correction to the eigenvalues is:
\be \delta \lambda_j = -\frac{2i(\Gamma_1^{(i)}+\Gamma_N^{(o)})}{N+1}\sin^2\frac{\pi j}{N+1}.\label{dlambda1}\ee
The correction is small in the thermodynamic limit $N\rightarrow\infty$. Unfortunately, the perturbation theory gives the correct result only for small values of $\Gamma_1^{(i)}$ and $\Gamma_N^{(o)}$, as the $l$-th order corrections in the perturbation theory with respect to $\Gamma_1^{(i)}$ and $\Gamma_N^{(o)}$ give corrections in the leading order with respect to $1/N$ proportional to $(\Gamma_1^{(i)}+\Gamma_N^{(o)})^l/N$. 
Fortunately, however, for an open system of non-interacting fermions we can write explicitly the characteristic polynomial of the matrix $\mathcal{N}$:
\bea C(\Lambda)& =& \frac{-y^N z^2 (z - 2 i (\Gamma_1^{(i)} - \Lambda)) (2i + 
      y (\Gamma_N^{(o)} - \Lambda)) + 
   y^2 z^N (y - 2 i (\Gamma_1^{(i)} - \Lambda)) (2 i + 
      z (\Gamma_N^{(o)} - \Lambda))}{2^N y^2 z^2 \sqrt{-4-\Lambda^2}},\\
       &&y=-i\Lambda - \sqrt{-4-\Lambda^2},z=-i\Lambda + \sqrt{-4-\Lambda^2}, \Lambda= i \lambda.\eea 
(here $t=1$). 
$C(\Lambda)$ has its zeros at $\Lambda_j=i \lambda^{(0)}_j$ for $\Gamma_1^{(i)}=0$ and $\Gamma_N^{(o)}=0$. The Taylor expansion gives the value of the correction of $\Lambda$:
\bea && C(i \lambda^{(0)}_j + \delta \Lambda,\Gamma_1^{(i)},\Gamma_N^{(o)})=0,\\
 &0&=C(i \lambda^{(0)}_j,\Gamma_1^{(i)},\Gamma_N^{(o)})+\delta\Lambda  C'_{\Lambda}(i \lambda^{(0)}_j ,\Gamma_1^{(i)},\Gamma_N^{(o)}).\eea
The straightforward calculation, which takes into account the functional form of $\lambda^{(0)}_j$ and the expansion for $\psi_j=\frac{\pi j}{N+1}\ll 1$, leads us to
\be \delta \lambda_j = -i \frac{2 \psi_j^2}{N}\left(\frac{\Gamma_1^{(i)}}{1+\Gamma_1^{2(i)}}+\frac{\Gamma_N^{(o)}}{1+\Gamma_N^{2(o)}} \right) - \frac{2\psi_j^2}{N}\left(\frac{\Gamma_1^{2(i)}}{1+\Gamma_1^{2(i)}}+\frac{\Gamma_N^{2(o)}}{1+\Gamma_N^{2(o)}} \right), 1\ll N \label{deltaL}.\ee
In first order in $\Gamma_{1,N}^{(i,o)}$ the expansion of the  imaginary part of (\ref{deltaL}) coincides with the previously derived expression~(\ref{dlambda1}) for  $\psi_j\ll 1$. The real part of the correction has not been explicitly derived above, as it is only of the second-order in $\Gamma_{1,N}^{(i,o)}$, so it was not derived in the first-order perturbation theory discussed above. Of course it coincides with the result of the second order perturbation theory. (The first order correction with respect to  $\Gamma_{1,N}^{(i,o)}$ and $1/N$ for $\lambda$'s for the XX-chain was derived in Ref.~\onlinecite{Zni}). 
Comparing the obtained correction with the perturbation theory expansion we can rewrite (\ref{deltaL}) as
\be \delta \lambda_j = -i \frac{2 \sin^2\psi_j}{N}\left(\frac{\Gamma_1^{(i)}}{1+\Gamma_1^{2(i)}}+\frac{\Gamma_N^{(o)}}{1+\Gamma_N^{2(o)}} \right) - \frac{2\sin^2\psi_j}{N}\left(\frac{\Gamma_1^{2(i)}}{1+\Gamma_1^{2(i)}}+\frac{\Gamma_N^{2(o)}}{1+\Gamma_N^{2(o)}} \right), 1\ll N \label{deltaL2_2} \nonumber\ee
as the all terms of order $1/N$ come in perturbation theory with the the prefactor $\sin^2\psi_j$.

\subsection{Eigenvectors}
\label{sec:EigenVectors}
We will also need the corrections to the eigenvectors. 
The eigenvectors of the matrix $\mathcal{M}$ which determines the transformation between the basis $\{a\}$ and $\{f\}$ should be determined from the secular equation~\cite{LL3} in the perturbation theory as at the absence of a coupling to an environment the eigenvalues are two-fold degenerate. Turning on couplings to reservoirs lifts this degeneracy and the eigenvectors for the eigenvalues with a negative imaginary part are 
$( \tfrac{1}{\sqrt{2}}\{\phi_j^{(0)}\}, -\tfrac{i}{\sqrt{2}}\{\phi_j^{(0)}\} )$ and for the 
eigenvalues with a positive imaginary part are $(\tfrac{1}{\alpha}\{\phi_j^{(0)}\}, 
i\tfrac{\Gamma_1^{(i)}}{\alpha\Gamma_2^{(0)}}\{\phi_j^{(0)}\})$, 
$\alpha=\sqrt{1+\left(\tfrac{\Gamma_1^{(i)}}{\Gamma_2^{(o)}}\right)^2}$, for the 
infinitesimally small $\Gamma_1^{(i)}$ and $\Gamma_2^{(o)}$.

It is possible to notice that for the eigenvectors corresponding to the eigenvalues with a negative imaginary part the structure $(\tfrac{1}{\sqrt{2}}\{\phi_j\}, -\tfrac{i}{\sqrt{2}}\{\phi_j\})$ holds for any coupling to the reservoirs. Therefore, the first order correction to these eigenvectors can be calculated from the simple-minded first-order perturbation theory for the matrix $\mathcal{N}$:
\be \delta \phi_{k,j} = i \frac{1}{N+1}\sum_{m, m\neq k} \frac{(\Gamma_1^{(i)}+(-1)^{m+k}\Gamma_N^{(o)})\sin \frac{\pi m}{N+1}\sin \frac{\pi k}{N+1}}{ (\cos \frac{\pi m}{N+1} -\cos \frac{\pi k}{N+1})}\phi^{(0)}_{m,j}. \label{evcorr}\ee

In the limit of large $N$, $1\ll N$, the summation over $m$ is performed according to:
\bea \sum_{m, m\neq k} \frac{ \sin \frac{\pi m}{N+1} \sin \frac{\pi m j}{N+1}}{(\cos \frac{\pi m}{N+1} -\cos \frac{\pi k}{N+1})} &=& -(N+1) \left(1-\frac{j}{N+1} \right)\cos \frac{\pi j k}{N+1} + \cot \frac{\pi k}{N+1} \sin \frac{\pi j k}{N+1}, \\
\sum_{m, m\neq k} (-1)^m\frac{ \sin \frac{\pi m}{N+1} \sin \frac{\pi m j}{N+1}}{(\cos \frac{\pi m}{N+1} -\cos \frac{\pi k}{N+1})} &\approx & (-1)^k \left[j\cos \frac{\pi j k}{N+1} + \cot \frac{\pi k}{N+1} \sin \frac{\pi j k}{N+1}\right].
\eea
This leads for the correction to the eigenvector of $\mathcal{N}$:
\be \delta \phi_{k,j}^{\im \lambda>0} = 
-\frac{\sqrt{2}i}{(N+1)^{3/2}} (N+1) \sin \frac{\pi k }{N+1} \cos \frac{\pi j k }{N+1} \left[ \Gamma_1^{(i)}  \left( 1 - \frac{j}{N+1}\right)- \Gamma_N^{(o)}  \frac{j}{N+1}\right].\label{evcorr2}\ee
The result~(\ref{evcorr2}) with the substitution $\Gamma_{1,N}^{(i,o)}\rightarrow\frac{\Gamma_{1,N}^{(i,o)}}{1+\Gamma_{1,N}^{2(i,o)}}$ describes the eigenvectors with good precision,~Fig.~\ref{eigenVect}A,B.  

It is also possible to do the calculation of the correction to these eigenvectors using the degenerate perturbation theory~\cite{LL3}. For the problem at hand, the first order of the degenerate perturbation theory does not preserve the symmetry $(\tfrac{1}{\sqrt{2}}\{\phi_j\}, -\tfrac{i}{\sqrt{2}}\{\phi_j\})$. Apparently the summation should be done to higher orders of the perturbation theory to restore the exact symmetry of the eigenvectors. 

For the eigenvectors which correspond to the eigenvalues with positive imaginary part, we could not find a matrix of the smaller size, for which the eigenvectors could be found by the first-order perturbation theory. Therefore, we can perform only the degenerate perturbation theory for these eigenvectors. Taking into account the comparison between the degenerate perturbation theory and the answer which describes the eigenvectors well, we have thrown away one term from the degenerate perturbation theory (which describes the correction due to the paired-degenerate level), and concluded that the correction to these eigenvectors can be described by the expression similar to~(\ref{evcorr2}), Fig.~\ref{eigenVect}C,D:
\be \label{evcorrForm}\delta \phi_{k,j}^{\im \lambda>0}(f_1,f_2) = \frac{i\sqrt{2}}{N^{1/2}} \sin \frac{\pi k }{N+1} \cos \frac{\pi j k }{N+1} \left[ f_1(\Gamma_1^{(i)},\Gamma_N^{(o)})  \left( 1 - \frac{j}{N+1}\right)- f_2(\Gamma_1^{(i)},\Gamma_N^{(o)})  \frac{ j }{N+1}
\right].\ee
The coefficients $f_{1,2}$ in this expression can not be found precisely from the first order degenerate perturbation theory. We would like to argue that they are linear in $\Gamma_{1,N}^{(i,o)}\rightarrow\frac{\Gamma_{1,N}^{(i,o)}}{1+\Gamma_{1,N}^{2(i,o)}}$.    

 Coefficients $f_1$ and $f_2$ in the expression~(\ref{evcorrForm}) can be found from the direct substitution into the matrix equation and requiring that the precision of the found eigenvectors is $O(1/N^{3/2})$ in the limit $N\rightarrow \infty$: 
\be (\mathcal{M}-\lambda_j I_{2N}) \begin{pmatrix} \tfrac{1}{\alpha}\{\phi_j^{(0)}+\delta \phi_j(b_1,b_2)\} \\ 
i\tfrac{\Gamma_1^{(i)}}{\alpha\Gamma_N^{(o)}}\{\phi_j^{(0)} +\delta \phi_j(a_1,a_2)\} 
\end{pmatrix} = O(1/N^{3/2}).
\ee
The requirements of the $O(1/N^{3/2})$ precision of the eigenvectors leads to:
\bea 
b_1 &=&\frac{\Gamma_1^{(i)}-i \Gamma_1^{2(i)}\cos \frac{\pi k }{N+1}} {1+\Gamma_1^{2(i)}\cos^2 \frac{\pi k }{N+1}}  \label{b1},\\ 
b_2&=& -\frac{2\Gamma_1^{(i)}\cdot(-1)^{k+1}+\Gamma_N^{(o)}+i \Gamma_N^{2(o)} \cos\frac{\pi k }{N+1}} {1+\Gamma_2^{2(o)} \cos^2 \frac{\pi k }{N+1}},\\
a_1&=& \frac{\Gamma_N^{(o)}-i \Gamma_N^{2(o)} \cos\frac{\pi k }{N+1}} {1+\Gamma_N^{2(o)}\cos^2 \frac{\pi k }{N+1} },\label{a1}\\
a_2 &=&-\frac{2\Gamma_1^{(i)}+4\cdot (-1)^{k+1}\cdot\Gamma_N^{(o)}+i \Gamma_1^{2(i)}\cos \frac{\pi k }{N+1}} {2+\Gamma_1^{2(i)}(1+\cos\frac{2\pi k }{N+1})}  .\label{a2}
\eea
Let us notice that all the coefficients multiplied by the $k$ dependent term $\cos \frac{\pi k }{N+1}$ are of the second order in 
$\Gamma$. Therefore, for small values of $\Gamma$ we can neglect $k$-dependence of these coefficients.

We have also checked the validity of the procedure of the substitution of the form of the eigenvectors and finding the relevant coefficients for the case of the eigenvectors corresponding to the eigenvalues with negative imaginary part and have found that the coefficients $b_1$ is the same as Eq.~(\ref{b1}) and $b_2$ is opposite in sign to Eq.~(\ref{a1}) (the coefficient for the lower part are found from the symmetry requirement of the structure of the eigenvectors).
It is consistent with the first order calculation and an insight that the couplings $\Gamma_1^{(i)}$ and $\Gamma_2^{(o)}$ should enter in the similar combination as in the correction $\delta\lambda_j$, Eq.~(\ref{deltaL2_2}).

There is one case where we can recover the symmetry of the eigenvectors corresponding to eigenvalues with a positive imaginary part. It is $\Gamma_1^{(i)}=\Gamma_N^{(o)}=\Gamma$. Then the eigenvectors have the symmetry $(\tfrac{1}{\sqrt{2}}\{\phi_{2p}\}, i\tfrac{1}{\sqrt{2}}\{\phi_{2p}\}^{T})$ or $(\tfrac{1}{\sqrt{2}}\{\phi_{2p+1}\}, -i\tfrac{1}{\sqrt{2}}\{\phi_{2p+1}\}^{T})$. Therefore, they can be found as eigenvectors corresponding to eigenvalues with positive imaginary part of the matrices, correspondingly: 
\be \label{Nprime} \mathcal{N}^\prime = \begin{pmatrix}[ccccc]
                        -i\Gamma & t & 0 & \ldots& 0  \cr
                        t & 0 & t & \ddots & \vdots  \cr
                        0 & \ddots & \ddots & \ddots & \vdots  \cr                
                        \vdots &  \ldots &\ddots & 0 & t \cr
                       
                        2 i \Gamma & \ldots & 0 & t & -i\Gamma
                       \end{pmatrix},~~~
                       \mathcal{N}^{\prime\prime} = \begin{pmatrix}[ccccc]
                        -i\Gamma & t & 0 & \ldots& 0  \cr
                        t & 0 & t & \ddots & \vdots  \cr
                        0 & \ddots & \ddots & \ddots & \vdots  \cr                
                        \vdots &  \ldots &\ddots & 0 & t \cr
                       
                         -2 i \Gamma & \ldots & 0 & t & -i\Gamma
                       \end{pmatrix}
\ee
The eigenvectors after the first order calculation look similar to Eq.~(\ref{evcorr}):
\be \delta \phi_{k,j} = - i \frac{\sqrt{2}\Gamma}{(N+1)^{1/2}}\sum_{m, m\neq k} \frac{(1+3\cdot(-1)^{m+k})\sin \frac{\pi m}{N+1}\sin \frac{\pi k}{N+1}}{(\cos \frac{\pi m}{N+1} -\cos \frac{\pi k}{N+1})}\phi^{(0)}_{m,j}. \label{evcorrPos}\ee
The summation leads to Eq.~(\ref{evcorr2}), with coefficients consistent with expressions~Eqs.(\ref{b1})-(\ref{a2}).
The answer for larger $\Gamma$ is obtained by substitution 
$\Gamma\rightarrow\frac{\Gamma}{1+\Gamma^{2}}$, Fig.~\ref{eigenVect}E,G.  

\begin{figure}[tb!]
\begin{center}
A\includegraphics[width=0.25\linewidth]{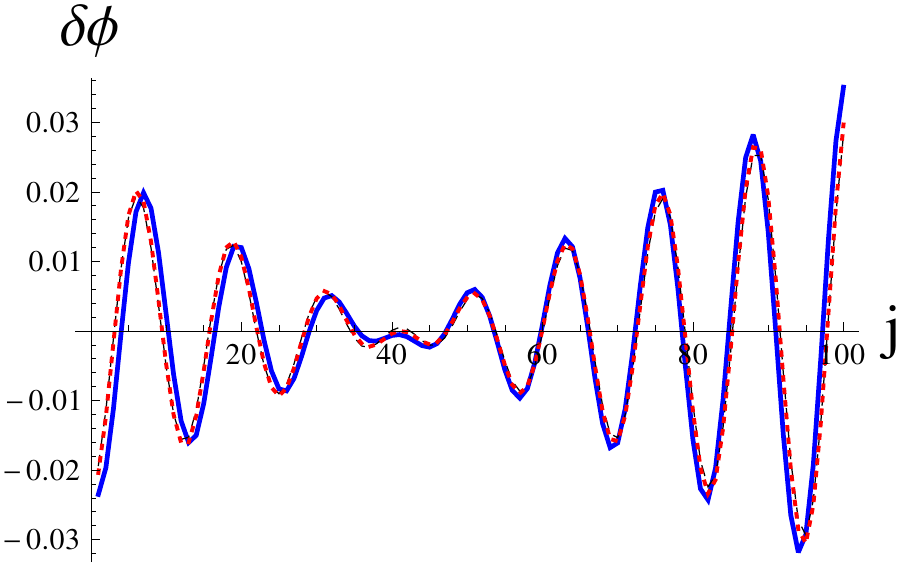}
B\includegraphics[width=0.25\linewidth]{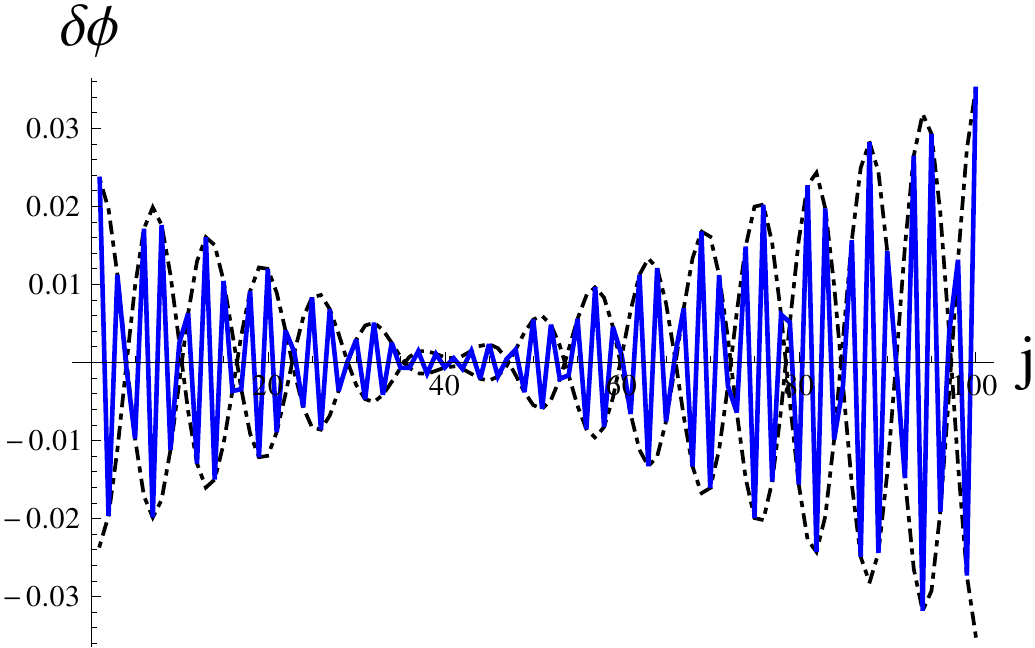}
C\includegraphics[width=0.25\linewidth]{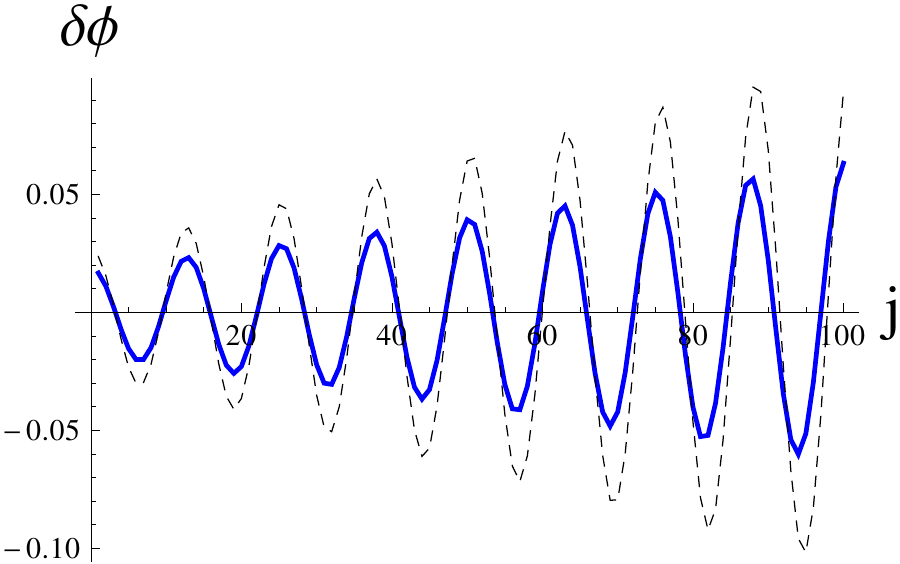}
D\includegraphics[width=0.25\linewidth]{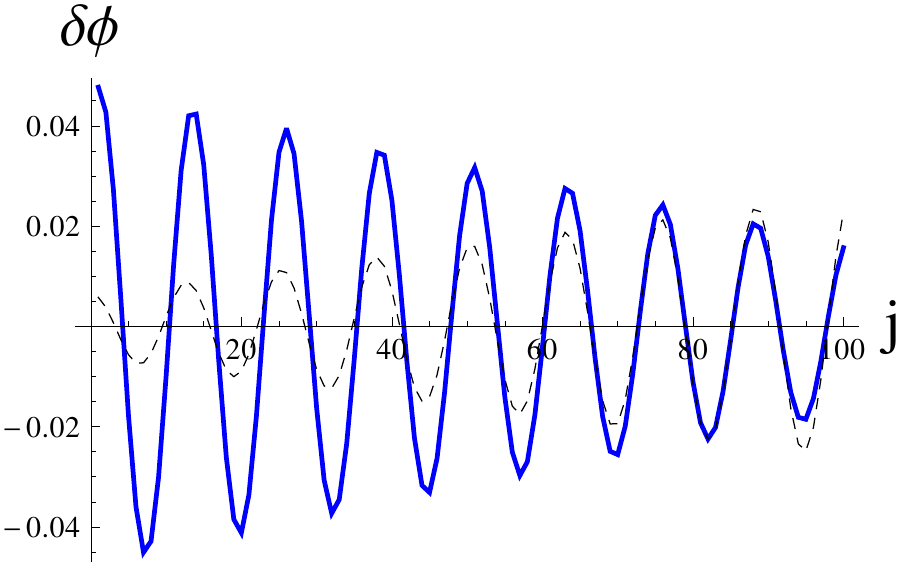}
E\includegraphics[width=0.25\linewidth]{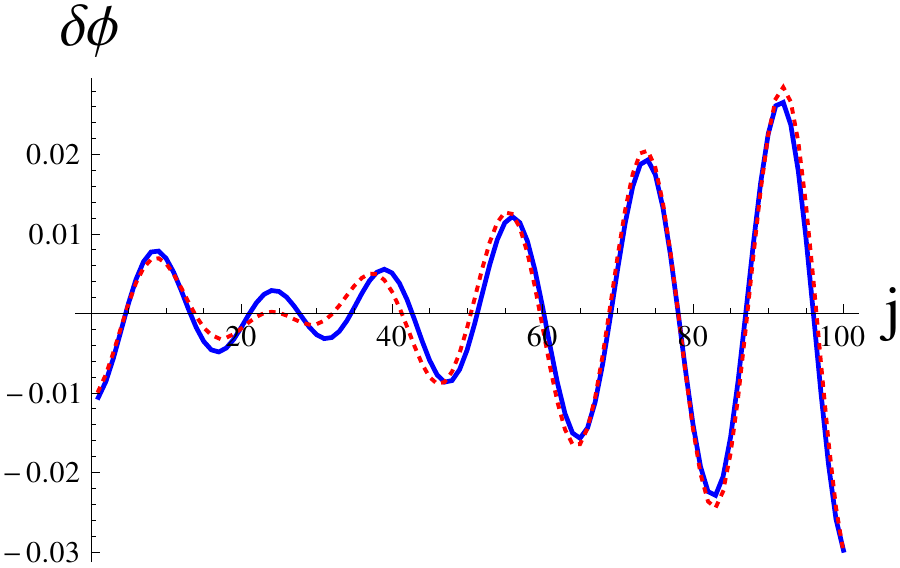}
G\includegraphics[width=0.25\linewidth]{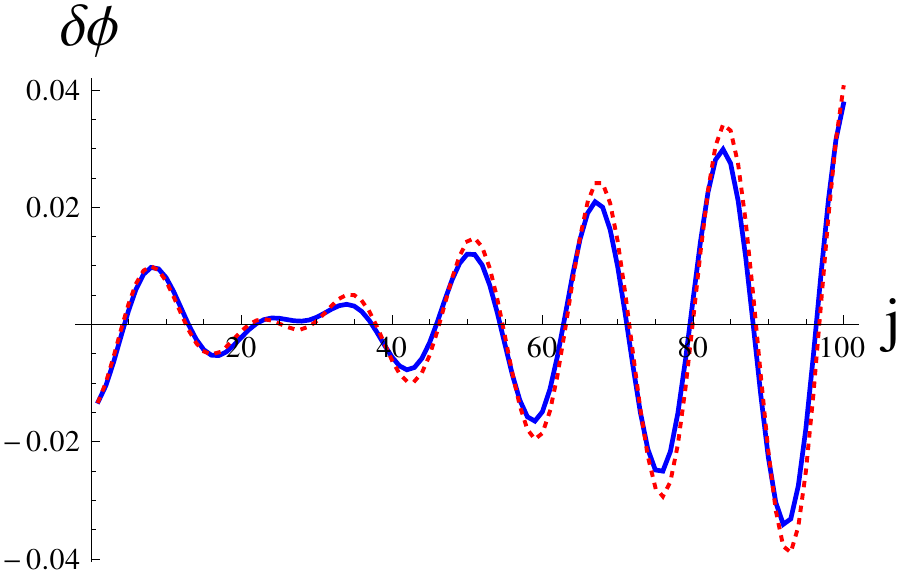}

\caption{\label{eigenVect}
The correction to the imaginary part of the eigenvectors. Parameters: $\Gamma_{1}^{(i)}=2.6,\Gamma_{N}^{(o)}=1.3$. Blue solid line: imaginary part of the eigenvector from the numerical calculation, black dashed line: the perturbative correction~(\ref{evcorr}) taking into account the substitution $\Gamma_{1,N}^{(i,o)}\rightarrow\frac{\Gamma_{1,N}^{(i,o)}}{1+\Gamma_{1,N}^{2(i,o)}}$, red dotted line: the approximation of the sum ~(\ref{evcorr2}). 
A: Correction to the eigenvector corresponding to the eigenvalue with negative imaginary part for $k\le N/2$ and B: for $k > N/2$, dashed-dotted lines represent the envelope.  
C, D, E: Correction to the eigenvector corresponding to the eigenvalue with positive imaginary part, elements $1$ to $N$ (C), elements $N+1$ to $2N$ (D).
E,G: is the case of $\Gamma_1^{(i)}=\Gamma_N^{(o)}=4.3,~0.3$. 
}\end{center}
\end{figure}

\subsection{Inverting the matrix of the eigenvectors}
\label{subsec:Invertion}

The matrix of the eigenvectors can be written in the form: 
\be \label{P} P = \begin{pmatrix}[c|c]
\tfrac{1}{\sqrt{2}}\phi + \delta \phi_1 & \tfrac{1}{\alpha}\phi+\delta\phi_2 \\ 
\hline
-i\tfrac{1}{\sqrt{2}}(\phi + \delta \phi_1) & \tfrac{1}{\alpha}\tfrac{\Gamma_{N}^{(o)}}{\Gamma_{1}^{(i)}} \phi+\delta\phi_3 
\end{pmatrix},
\ee
where all corrections are of the order $\tfrac{1}{\sqrt{N}}\tfrac{\Gamma_{1,N}^{(i,o)}}{1+\Gamma_{1,N}^{2(i,o)}}$,  $\delta \phi_1$ is (\ref{evcorr2}), all of the corrections $\delta\phi_{1,2,3}$ have the form~(\ref{evcorrForm}). The inverse of matrix~(\ref{P}) has  structure: 
\be \label{T} T = \begin{pmatrix}[c|c]
\tfrac{\sqrt{2}}{1+\Gamma_{N}^{(o)}/\Gamma_{1}^{(i)}}\phi + l_1(\delta \phi) & \tfrac{\sqrt{2}i}{1+\Gamma_{1}^{(i)}/\Gamma_{N}^{(o)}} \phi+ l_3(\delta\phi) \\ 
\hline
\tfrac{\alpha}{1+\Gamma_{1}^{(i)}/\Gamma_{N}^{(o)}} \phi + l_2(\delta \phi) & \tfrac{-i\alpha}{1+\Gamma_{1}^{(i)}/\Gamma_{N}^{(o)}}\phi+ l_4(\delta\phi) 
\end{pmatrix},
\ee
where $l_i(\delta\phi)$ are the transformations of the  linear combinations of the corrections $\delta \phi_{1,2,3}$.
This transformation is determined using a standard expression for the inverse of the block matrix. Taking into account that for the matrix of the unperturbed eigenvalues $\{\phi \} = \{\phi \}^{-1}$ neglecting the $k$-dependence of the coefficients $a_1,a_2,b_1,b_2$, Eqs.(\ref{b1})-(\ref{a2}), and assuming that the correction to the matrix of unperturbed  eigenvectors is small (they are small for $\tfrac{\Gamma_{1,N}^{(i,o)}}{1+\Gamma_{1,N}^{2(i,o)}}\ll 1$), we get the correction for matrix $T$ in the form:
\be \lim_{N \rightarrow \infty}l_s(\delta\phi)_{km} = \frac{1}{(N+1)^{3/2}}\sum_{i=1,j=1}^{i=N,j=N} \sin\frac{\pi i m }{N+1} \cos\frac{\pi i j }{N+1} \sin\frac{\pi i}{N+1} \left( \gamma_1  + \gamma_2\frac{j}{N+1} \right)\sin\frac{\pi j k }{N+1},\ee
where the coefficients $\gamma_1$ and $\gamma_2$ stand for the coefficients $f_1$ and $f_2$ in the expression of the form Eq.~(\ref{evcorrForm}). 

 Let us perform the summation over index $i$ first:
\bea \sum_{i=1}^{N} \sin\frac{\pi i m }{N+1} \cos\frac{\pi i j }{N+1} \sin\frac{\pi i}{N+1} = \frac{N+1}{4} \left(-\delta_{m,j-1} +\delta_{m,j+1}\right). 
\eea
Now the summation over index $j$ is trivial and gives:
\be \label{CorrInverse} \lim_{N \rightarrow \infty}l_s(\delta\phi)_{km} = -\frac{1}{2(N+1)^{1/2}} \left[ \left(\gamma_1 + \gamma_2 \frac{m}{N+1}\right) \cos\frac{\pi k m }{N+1}\sin\frac{\pi k}{N+1} + \frac{\gamma_2}{N+1}  \cos\frac{\pi k}{N+1}\sin\frac{\pi k m}{N+1}    \right]. \ee 
The expression above has the same function form as corrections to the eigenvalues, Eq.~(\ref{evcorrForm}), neglecting the term proportional to $\frac{\gamma_2}{N+1}$ which is small in the limit of large $N$.

We are interested in the transpose of the matrix $T$ to determine the coefficients $C^{(1)}_{mk}$. They are given by  $l_s(\delta\phi)_{mk}$.

The current is given by $ C^{(1)}_{m j} A^{(2)}_{nj}$  and
$C^{(1)}_{m k} A^{(2)}_{nj}$. These are (\ref{cor2}) and (\ref{cor22}) from the main text
close to the ends of the chain and for small $k$ ($\frac{k n \pi}{N+1}\ll 1$) in the thermodynamic limit.

Far from the ends of the chain we are not allowed to expand the trigonometric functions of $\frac{k n \pi}{N+1}$ in the Taylor series. It suggests the possible deviation of the corrections Eqs.~(\ref{cor2}),~(\ref{cor22}) far from the ends of the chain. We do not discuss these deviations in this work and leave it for further investigation.  
\end{widetext}

\end{document}